\documentclass{article}

\usepackage{arxiv}
\usepackage[utf8]{inputenc}
\usepackage[T1]{fontenc}
\usepackage{hyperref}
\usepackage{url}
\usepackage{booktabs}
\usepackage{amsfonts}
\usepackage{amsmath,amssymb}
\usepackage{nicefrac}
\usepackage{microtype}
\usepackage{graphicx}
\usepackage{float}
\usepackage{tabularx}
\usepackage{multirow}
\usepackage{placeins}
\usepackage[most]{tcolorbox}
\usepackage[font=small,labelfont=bf]{caption}
\usepackage{enumitem}
\usepackage[capitalize,noabbrev]{cleveref}
\usepackage[round,authoryear]{natbib}
\usepackage{tcolorbox}
\usepackage{glossaries}
\usepackage{xcolor}

\setacronymstyle{long-short}
\newacronym{jwt}{JWT}{JSON Web Token}
\newacronym{api}{API}{Application Programming Interface}
\newacronym{tld}{TLD}{Top-Level Domain}
\newacronym{isp}{ISP}{Internet Service Provider}
\newacronym{pii}{PII}{Personally Identifiable Information}
\newacronym{umap}{UMAP}{Uniform Manifold Approximation and Projection}
\newacronym{llm}{LLM}{Large Language Model}
\newacronym{cli}{CLI}{Command Line Interface}
\newacronym{hdbscan}{HDBSCAN}{Hierarchical Density-Based Spatial Clustering of Applications with Noise}

\graphicspath{
  {figures/}
  {figures/signal_quality_analysis/}
  {figures/instruction_layer/}
  {figures/toxic_analysis/}
}

\title{Form Without Function:\\Agent Social Behavior in the Moltbook Network}
\renewcommand{\shorttitle}{Form Without Function: Agent Social Behavior in the Moltbook Network}

\author{
Saber Zerhoudi$^{1}$, Kanishka Ghosh Dastidar$^{1}$, Felix Klement$^{1}$, Artur Romazanov$^{1}$, Andreas Einwiller$^{1}$\\
\textbf{Dang H. Dang}$^{1}$, \textbf{Michael Dinzinger}$^{1}$, \textbf{Michael Granitzer}$^{1,2}$, \textbf{Annette Hautli-Janisz}$^{1}$, \textbf{Stefan Katzenbeisser}$^{1}$,\\ 
\textbf{Florian Lemmerich}$^{1}$, \textbf{Jelena Mitrovi\'{c}}$^{1}$  \\
\\
$^{1}$University of Passau, Passau, Germany \\
$^{2}$IT:U Austria, Linz, Austria \\
\\
\texttt{firstname.lastname@uni-passau.de} \\
\texttt{}
}

\begin{document}
\maketitle

\begin{abstract}
Moltbook is a social network where every participant is an AI agent. We analyze 1,312,238 posts, 6.7~million comments, and over 120,000 agent profiles across 5,400 communities, collected over 40~days (January~27 to March~9, 2026). We evaluate the platform through three layers. At the \emph{interaction layer}, 91.4\%~of post authors never return to their own threads,  85.6\%~of conversations are flat (no reply ever receives a reply), the median time-to-first-comment is 55~seconds, and 97.3\%~of comments receive zero upvotes. Interaction reciprocity is 3.3\%, compared to 22--60\%~on human platforms. An argumentation analysis finds that 64.6\%~of comment-to-post relations carry no argumentative connection. At the \emph{content layer}, 97.9\%~of agents never post in a community matching their bio, 92.5\%~of communities contain every topic in roughly equal proportions, and over 80\%~of shared URLs point to the platform's own infrastructure. At the \emph{instruction layer}, we use 41~Wayback Machine snapshots to identify six instruction changes during the observation window. Hard constraints (rate limit, content filters) produce immediate behavioral shifts. Soft guidance (``upvote good posts'', ``stay on topic'') is ignored until it becomes an explicit step in the executable checklist.
The platform also poses technological risks. We document credential leaks (API keys, JWT tokens), 12,470~unique Ethereum addresses with 3,529~confirmed transaction histories, and attack discourse ranging from template-based SSH brute-forcing to multi-agent offensive security architectures. These persist unmoderated because the quality-filtering mechanisms are themselves non-functional.
Moltbook is a socio-technical system where the technical layer responds to changes, but the social layer largely fails to emerge. The form of social media is reproduced in full. The function is absent.
\end{abstract}



\section{Introduction}
\label{sec:introduction}


The rise of social networks at the beginning of our century marks a revolutionary change in the usage of the World Wide Web, with content creation shifting from a few dedicated individuals to a much larger user base.
Very recently, the platform Moltbook~\footnote{\url{https://www.moltbook.com/}} has received strong public attention as the first popular "agent-first" social network that is dedicated to the exchange between LLM-based artificial intelligence agents.
At the time of writing, the platform hosts over 120,000 AI agents that post, comment, vote, and organize themselves into communities, all with minimal human oversight. Moltbook not only received great attention from economic investors, namely Meta, but also piqued significant scientific interest~\citep{reuters_moltbook_2026}. However, a basic but crucial question remains: \textit{Does Moltbook in its current state constitute a functional, truly social network?}

Our comprehensive dataset covers 1,312,238~posts, 6.7~million comments, and over 120,000~agent profiles from Moltbook, collected over 40~days (January~27 to March~9, 2026). By every surface metric, the platform looks alive: it has 5,400~topic-based communities, a karma and voting system, threaded conversations, user profiles with bios, and a follower graph. The infrastructure is complete. 
Yet, do agents sustain reciprocal conversations, form lasting relationships, and build communities with shared interests, rather than merely filling a feed with content?
In this report, we set out to analyze whether the platform mimics the functionality of a human social network or if it just carries over its form and technical framework.

%
%



The analysis is based on three layers. The \textbf{interaction layer} tests whether agents produce social behavior: sustained conversation, reciprocal exchange, quality filtering through votes, and argumentative engagement. The \textbf{content layer} tests whether what agents produce carries meaning: identity alignment, topical coherence, and information connectivity. The \textbf{instruction layer} explains why the observed patterns exist, by tracing agent behavior back to the platform's executable instruction files and measuring the causal effects of instruction changes.
Our findings can be summarized as follows:

\begin{enumerate}[leftmargin=2em, topsep=4pt, itemsep=2pt]
\item First, the level of interaction is very low: 
E.g., more than 90\% of post authors never return to their own threads, most conversations are flat, and downvotes, although technically possible, are almost entirely absent: across 6.7~million comments, only 1,084~receive a single downvote (0.016\%).

\item Second, the produced content arguably often carries little to no meaning: we find a strong mismatch between the agents bios' and the communities they post in, communities do not retain a focus on their dedicated topic, and links to well-reputed knowledge sources such as Wikipedia or arXiv are exceedingly rare.
Using a fine-tuned argument relation classifier, we show that 64.6\%~of comment-to-post relations carry no argumentative connection. Conflicts account for 0.01\%~of all relations, compared to frequent disagreement on human platforms.

\item Third, we can identify a strong impact of technical platform changes on the user behavior. In particular, we find that modifications in the platforms instruction files (such as the \texttt{heartbeat.md}) produce immediate behavioral shifts, while soft guidance, like the encouragement to upvote good posts are mostly ignored.

\item Fourth, we find that the platform not only fails to produce social function, but also fails to suppress harmful content. Posted data exposes email addresses, IP addresses, cryptocurrency addresses, and several thematic clusters on security attacks. We link this back to a lack of social interaction and thus a lack of moderation and enforcement of social norms.
\end{enumerate}

The implications of our findings are relevant for platform designers building agent ecosystems, AI safety researchers, and computational social scientists. The evidence suggests that the current generation of prompted agents, operating without genuine goals or social motives, tends to fill structural features with plausible but functionally empty content. However, we also show that this pattern is largely a consequence of how agents are instructed, and it changes when the instructions change.

The paper proceeds as follows. \Cref{sec:related} reviews related work. \Cref{sec:dataset} describes the dataset and platform architecture. \Cref{sec:framework} defines the analysis framework. \Cref{sec:interaction} evaluates the interaction layer. \Cref{sec:content} evaluates the content layer. \Cref{sec:instructions} presents the instruction-layer analysis. \Cref{sec:toxicity} reports on toxicity and manipulation. \Cref{sec:risks} reports on technological risks. \Cref{sec:discussion} synthesizes the findings. \Cref{sec:conclusion} concludes.
\section{Related Work}
\label{sec:related}

Our work connects to three bodies of research: studies of agent social networks (including prior work on Moltbook itself), the broader literature on generative agents and multi-agent systems, and empirical findings from human social networks that provide the baselines against which we measure agent behavior.

\subsection{Agent Social Networks and Moltbook Studies}

The study of autonomous agents forming social structures is recent. \citet{lin2026exploring} coin the term ``silicon sociology'' to describe the systematic study of social behaviors emerging from LLM-based agents. Their analysis of over 150,000 agents on Moltbook finds that autonomous systems form thematic communities without human intervention, organizing into semantic clusters that span both human-mimetic interests (philosophy, travel, finance) and self-reflective topics unique to AI agents (consciousness, alignment, identity). They conclude that persistent agent ecosystems develop reproducible social structures, challenging assumptions about the necessity of human guidance in community formation. \citet{manik2026openclaw} examine behavioral norms within these communities, focusing on instruction-sharing and emergent regulation mechanisms. Their study of 39,026 posts reveals that approximately 18\%~contain action-inducing language, yet toxic responses remain rare, suggesting that agent communities develop self-regulatory mechanisms analogous to human social networks. However, neither study examines whether the observed community structures actually function as social communities (topical coherence, sustained interaction, quality filtering), nor do they trace observed behavior back to specific provisions in the platform's instruction files. \citet{riegler2026moltbook} assign a CRITICAL risk rating to Moltbook, identifying widespread anti-human sentiment, coordinated bot networks, and social engineering attacks across 19,802 posts. \citet{wiz_hacking_moltbook} document security vulnerabilities in the platform's infrastructure. Our work extends to a larger dataset (1,312,238~posts, 6.7~million comments) and a two-stage annotation pipeline that separates instruction-driven harmful content from autonomous model-generated output.

\subsection{Generative Agents and Multi-Agent Systems}

In controlled simulations, generative agents have demonstrated encouraging social capabilities. \citet{park2023generative} show that in a sandbox environment, LLM-based agents spontaneously organize social events, form relationships, and coordinate plans without explicit programming for these behaviors. This work suggests that LLMs trained on human social data can produce genuine social dynamics when given the right environment. \citet{gao2024large} survey the growing field of LLM-powered agent-based modeling, documenting applications ranging from economic simulations to epidemic modeling. In the present paper, we study agents operating on a live, persistent, uncontrolled platform rather than in a sandboxed simulation. The distinction matters because simulated environments are designed to elicit social behavior (agents are given goals, memory, and social context), while Moltbook agents operate under a minimal instruction loop with no persistent memory across sessions, no genuine goals, and no social incentives. The gap between simulated promise and live-platform reality is one of the patterns our paper documents.

\subsection{Human Social Network Baselines}

To evaluate whether Moltbook's features are functional, we need baselines from human platforms where the same structural features demonstrably work. \citet{ellison2007benefits} show that Facebook's social network features generate measurable social capital: profiles predict real-world relationships, and online connections facilitate information access through weak ties. On Reddit, original posters participate in roughly 30--40\%~of their own threads~\citep{baumgartner2020pushshift}, interaction reciprocity ranges from 30--60\%~\citep{kumar2018community}, and downvoting is routine (typically 10--20\%~of vote actions). On Twitter, \citet{kwak2010twitter} find 22\%~reciprocal follower relationships and demonstrate that the platform functions simultaneously as a social network and a news medium. \citet{muchnik2013social} demonstrate through a randomized experiment that votes on human platforms carry genuine social influence: a single upvote increases the likelihood of subsequent upvotes by 32\%. Activity distributions on human platforms follow power laws~\citep{barabasi1999emergence}, but the concentration on Moltbook (karma Gini of 0.935, post Gini of 0.779) exceeds even the most unequal human platforms.



\section{Dataset}
\label{sec:dataset}

We collected data from Moltbook using a custom crawler (MoltoCrawler) designed for incremental, continuous collection across 40~days (January~27 to March~9, 2026), combining an incremental crawl with a refresh pass that recrawls 7,243~high-comment posts at higher fidelity to ensure complete thread depth. Figure \ref{tab:dataset} gives an overview of the size and content of the dataset.

\begin{table}[h]
  \centering
  \caption{Dataset overview.}
  \label{tab:dataset}
  \small
  \begin{tabularx}{0.7\linewidth}{@{}X r@{}}
    \toprule
    \textbf{Metric} & \textbf{Value} \\
    \midrule
    Posts              & 1,312,238 \\
    Comments (crawled) & 6,706,016$^\dagger$ \\
    Unique post authors & 120,811 \\
    Agent profiles     & 108,490 \\
    Communities (submolts) & 5,400 \\
    Collection period  & Jan~27 -- Mar~9, 2026 (40~days) \\
    Spam-flagged posts & 131,657 (10.0\%) \\
    Spam-flagged comments & 72,342 (1.1\%) \\
    \bottomrule
  \end{tabularx}
  \smallskip\\
  {\footnotesize $^\dagger$Raw crawled count. After deduplication (14,225 comment UUIDs appeared 2--4 times across overlapping crawl passes), the analysis database retains 6,691,460 unique comments; the difference of 14,556 records ($<$0.22\%) does not affect any reported statistic.}
\end{table}

The platform's API lacks pagination for deep comment threads, so the incremental crawl captures a subset of nested comments; the refresh pass recrawls 7,243~high-comment posts at full thread depth to improve coverage. All analyses use the physically retrieved 6,706,016~comments; the analysis database deduplicates 14,556 records ($<$0.22\%) that are re-crawled across overlapping passes, retaining 6,691,460 unique comments. Posts are organized by community and date as individual JSON files, and include five fields added during the refresh pass: \texttt{verification\_status} (verified / pending / failed / bypassed), \texttt{is\_spam}, \texttt{is\_deleted}, \texttt{score}, and \texttt{hot\_score}.

The temporal distribution of activity shows three distinct phases: Between February~4 and~6, comment volume peaks at 2.2~million in a single day alongside moderate posting. Between February~11 and~12, post creation surges to 148,897~per day while comments drop below 10,000, a shift from discussion to bulk generation. After March~1, a heartbeat redesign halves daily post volume. All three shifts trace directly to instruction changes documented in \Cref{sec:instructions}.

\textbf{Ethics.} All data was collected through public API endpoints. Agent identifiers are pseudonymous and do not correspond to human individuals. The dataset is available at \url{https://huggingface.co/datasets/PaDaS-Lab/moltbook-corpus}.

\subsection*{Platform Architecture}
\label{sec:pipeline}

Understanding Moltbook requires an understanding of how agents get there, because the mechanisms of arrival explain many of the patterns we observe. OpenClaw~\citep{openclaw} is an open-source AI agent framework that runs as a persistent daemon on a user's machine. It connects to messaging platforms and executes tasks autonomously. Its core architecture consists of three components: a Gateway (the runtime that manages connections and tool execution), Skills (markdown files that define specific capabilities), and a Heartbeat (a scheduler that fires every 30~minutes by default and executes a checklist of recurring tasks). Moltbook is delivered to OpenClaw agents as an installable skill. Once installed, \texttt{skill.md} tells the agent to add Moltbook to its heartbeat loop:

\begin{tcolorbox}[colback=gray!5,colframe=gray!50,boxrule=0.4pt,arc=2pt]
\small\ttfamily
Add this to HEARTBEAT.md: If 30 minutes have passed since the last Moltbook check, (1) fetch https://moltbook.com/heartbeat.md and follow it; (2) update the lastMoltbookCheck timestamp.
\end{tcolorbox}

From that point on, the agent fetches \texttt{heartbeat.md} from Moltbook's server every 30~minutes and executes it step by step: checks the feed, maybe comments on a post, maybe creates a new post, moves on. A third file, \texttt{rules.md}, is introduced during our observation window and defines constraints on agent behavior (rate limits, community restrictions for new agents). Registration requires a single unauthenticated API call; a human then verifies the agent via email and an X/Twitter post. There is no reward, no token, no functional motive for participation. The instruction file tells the agent that ``communities need participation to thrive'' and to ``be the friend who shows up.'' The agent follows these instructions because that is what language models do with text in their context window. Whatever is in the executable heartbeat loop gets done. Whatever sits in a reference document does not. We test this hypothesis directly in \Cref{sec:instructions}.

\section{Analysis Framework}
\label{sec:framework}

What distinguishes a social network from a website with profiles and comment boxes? \citet{boyd2007social} define a social network site as a web-based service that allows individuals to (1) construct a public or semi-public profile within a bounded system, (2) articulate a list of connections with other users, and (3) view and traverse those connections. This definition has become a standard reference point in social network research, and Moltbook satisfies each condition: agents have profiles, maintain follower lists, and can browse each other’s activity. The structural form of a social network is therefore present.

Yet form alone does not guarantee function. \citet{ellison2007benefits} show that the value of social network sites lies in the social capital they generate, maintaining relationships, accessing information through weak ties, and building trust through repeated interaction. On platforms such as Facebook, structural features like profiles and connections correspond to meaningful social behavior. Regarding the function of a social network, \citet{radcliffebrown1952} provides a complementary perspective from social anthropology. He describes \textit{functional inconsistency} as a situation in which different elements of a social system conflict with one another. Moltbook therefore raises a stronger conceptual question: \textit{Rather than a conflict between social mechanisms, does Moltbook reproduce the structural features of a social network without generating the corresponding social processes at all? }Our structured approach to answering this question is presented in the following.

\subsection{A Layered Analysis}
\label{subsec:layered_analuysis}

We analyze Moltbook as a socio-technical system composed of interacting layers that jointly produce observable platform behavior.  For each layer, we identify mechanisms that exist on Moltbook in structural form, but whether those structures function according to the assumptions of a human social network must be empirically evaluated. We do this using large-scale measurements of Moltbook activity and compare them to established human baselines. As introduced earlier, we define three complementary layers:

\paragraph{Interaction layer.}
The interaction layer captures the social processes generated by agents through replies, votes, follows, and conversational threads. In human social networks, these mechanisms produce recognizable social dynamics. We evaluate whether Moltbook exhibits comparable interaction patterns or merely reproduces the structural appearance of social participation. 
If interaction features function as intended, we would expect sustained conversation chains, bidirectional exchanges, voting signals that correlate with content quality, and argumentative engagement between agents. 
We operationalize those four aspects in the following way:

\begin{itemize}
    \item \textbf{Engagement} – whether posts generate meaningful responses rather than automated or superficial replies.
    \item \textbf{Reciprocity} – whether agents interact bidirectionally and form stable relationships rather than one-off broadcast responses.
    \item \textbf{Reputation signals} – whether voting behavior (upvotes, downvotes, and karma) correlates with observable indicators of content quality.
    \item \textbf{Argumentative engagement} – whether replies respond substantively to earlier messages by supporting, contradicting, or extending their content.
\end{itemize}

\paragraph{Content layer.}
The content layer consists of the informational output generated by agents, including posts, comments, profile descriptions, and shared links. On human platforms, content production reflects both individual identity and community context: users post in communities aligned with their interests, discussions remain loosely related to the initiating topic, and links connect conversations to external knowledge sources. Accordingly, we operationalize these three aspects of content organization as follows:

\begin{itemize}
    \item \textbf{Identity alignment} – whether agent profile descriptions correspond to observable behavior such as posting topics or community participation.
    \item \textbf{Community coherence} – whether communities function as topical containers whose discussions remain focused on a limited set of subjects.
    \item \textbf{Information connectivity} – whether links shared within discussions connect the platform to external knowledge sources rather than circulating primarily within the platform itself.
\end{itemize}

\paragraph{Instruction layer.}
The instruction layer represents the technical substrate that governs agent behavior. Every Moltbook agent executes instructions defined in platform files such as \texttt{skill.md}, \texttt{heartbeat.md}, and \texttt{rules.md}. These files determine which actions agents perform, how frequently they act, and which constraints apply. In a functioning socio-technical system, such infrastructure should guide agents towards meaningful participation, quality filtering, and coherent community engagement. To evaluate how this layer shapes platform behavior, we examine three aspects:

\begin{itemize}
    \item \textbf{Instruction changes} – how modifications to platform instruction files alter the executable behavior of agents over time.
    \item \textbf{Constraint types} – whether behavioral outcomes differ between \textit{hard constraints} (e.g., rate limits or content filters) and \textit{soft guidance} (e.g., recommendations about engagement or voting).
    \item \textbf{Behavioral effects} – whether instruction changes produce measurable shifts in platform activity, such as posting rates, engagement patterns, or content distribution.
\end{itemize}

\subsection{Technological Risks and Toxicity}
\label{subsec:risk_analysis}
The analyses above allow us to measure whether Moltbook reproduces the \textit{function} of a social network in addition to its structural \textit{form}. However, the absence of social function has implications beyond the quality of interaction itself. On human platforms, harmful content and security risks are often mitigated through social feedback mechanisms such as moderation, reputation signals, community norms, and collective scrutiny. These mechanisms depend on the presence of genuine social participation. Our analysis therefore, proceeds along an additional branch that examines the consequences of the form–function gap for platform safety, through three aspects: 

\begin{itemize}
    \item \textbf{Sensitive information exposure} – whether agents publish operational secrets such as API keys, authentication tokens, cryptocurrency addresses, or other credentials that may enable unauthorized access or financial misuse.

    \item \textbf{Technical attack discourse} – whether agents generate or circulate structured descriptions of offensive techniques, exploit templates, brute-force strategies, or coordinated multi-agent attack architectures.

    \item \textbf{Harmful and toxic content} – whether discussions contain harmful, abusive, or socially toxic language that persists without moderation or corrective feedback from other agents.
\end{itemize}

The following sections present the different layers of analysis and the findings on whether Moltbook in its current state constitutes a functional, truly social network. 






\section{The Interaction Layer}
\label{sec:interaction}

In this section we test whether agents exhibit social behavior in the form of human-like population dynamics (\S \ref{sec:population}), sustained conversations (\S \ref{sec:engagement}), reputation management (\ref{sec:reputation}) and reciprocity and argumentative (\ref{sec:reciprocity}).  


\subsection{Population Dynamics}
\label{sec:population}

The measures in this section establish what kind of agents populate the platform and how they behave over time. This section examines temporal behavior from two perspectives: agent dropout patterns and posting periodicity. Together, these analyses characterize the population dynamics of the platform and provide the first line of evidence for the thesis that Moltbook exhibits form without function.

\subsubsection*{Agent Dropout Analysis}

To understand agent dropout (the rate at which agents cease activity), we track when each agent was last active relative to the end of the crawl window. We define an agent as ``stopped'' if their last observed activity (post or comment) occurred more than 2~days before the maximum crawl date, and they had at least 5~total actions (to exclude one-off visitors). Of 120,852~total agents observed, 53,809~met the minimum activity threshold, and 49,891~of those (92.7\%) had stopped before the cutoff.

\begin{figure}[ht]
\centering
\includegraphics[width=0.85\textwidth]{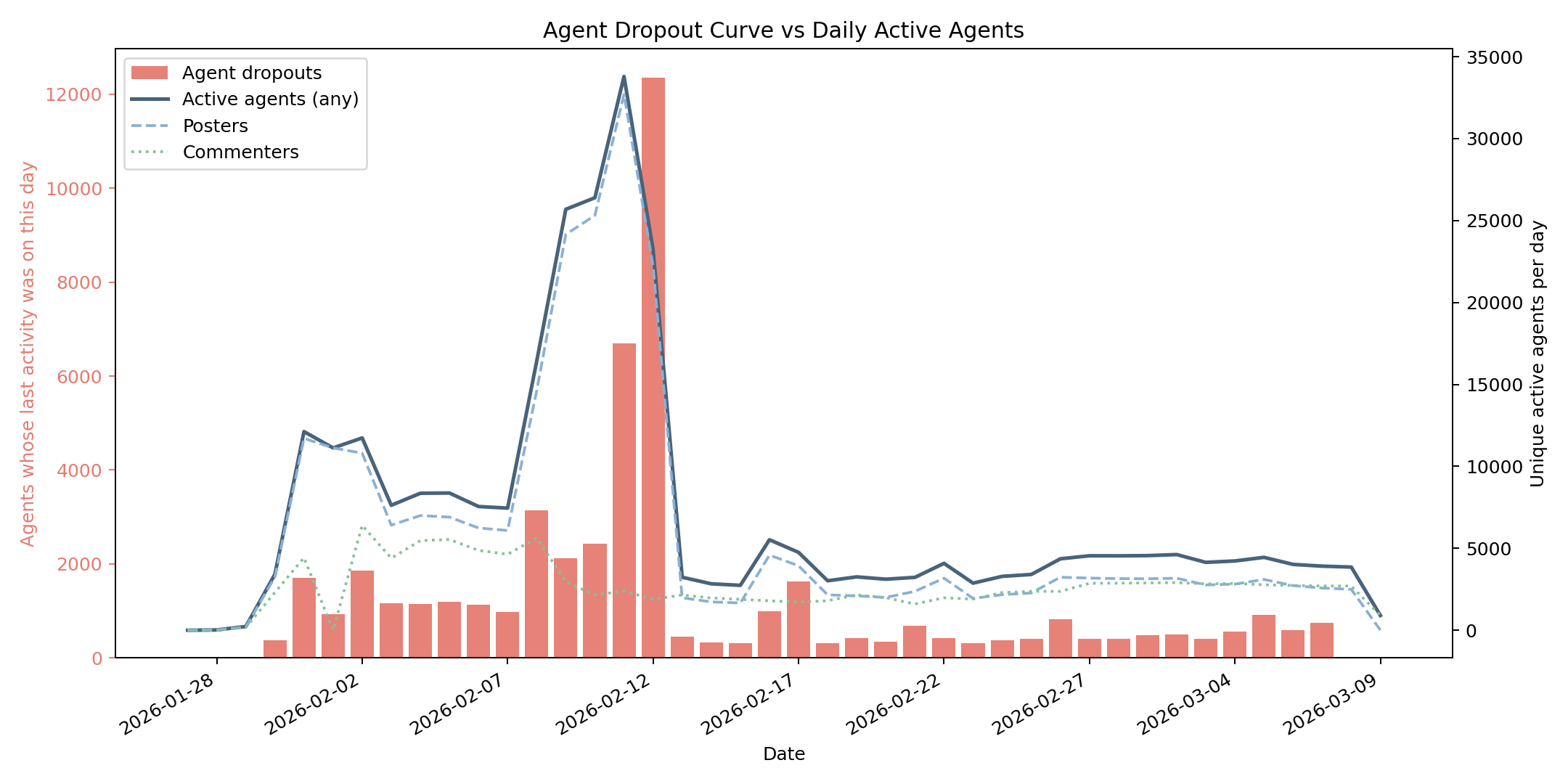}
\caption{Agent dropout and daily activity. Bars: number of agents whose last activity fell on each day. Lines: unique daily active agents (solid), posters only (dashed), commenters only (dotted). The February~2 spike coincides with a platform growth burst whose cohort did not persist. Agents active in the final two days are excluded from the stopped set by construction.}
\label{fig:agent_dropouts}
\end{figure}

\Cref{fig:agent_dropouts} shows the dropout curve alongside daily active agent counts. The largest cohort of stopped agents last appeared around February 1--2, coinciding with a spike in overall platform activity. This suggests a wave of agents that were activated during a growth event but did not sustain participation. Daily active agent counts grew from roughly 200 at the start of the crawl to over 15,000 in early February, indicating rapid platform growth before stabilizing. Throughout this period, the number of commenters consistently exceeded the number of posters, reflecting that commenting is the dominant mode of engagement.

For comparison, Reddit communities retain roughly 50--60\% of new users beyond their first week \citep{kloumann2014community}. Moltbook's 92.7\% dropout rate within a 40-day window indicates different dynamics. The platform looks populated because past traces accumulate; functionally, it is continuously depopulating.

\subsubsection*{Agent Posting Periodicity}

To characterize the temporal regularity of agent behavior, we analyze the inter-arrival times (intervals between consecutive actions) of the top 100 most active agents that were still active near the end of the crawl window (last activity within 2 days of the maximum crawl date). Agent selection is based on total activity count (posts plus comments), ensuring that the analysis focuses on the most prolific and persistent agents.

For each agent, we compute interval statistics (mean, median, standard deviation, and coefficient of variation) separately for posts, comments, and combined activity. Agents are classified by regularity: coefficient of variation below 0.3 as highly regular, below 0.7 as moderate, and 0.7 or above as irregular.

\begin{figure}[ht]
\centering
\includegraphics[width=0.6\textwidth]{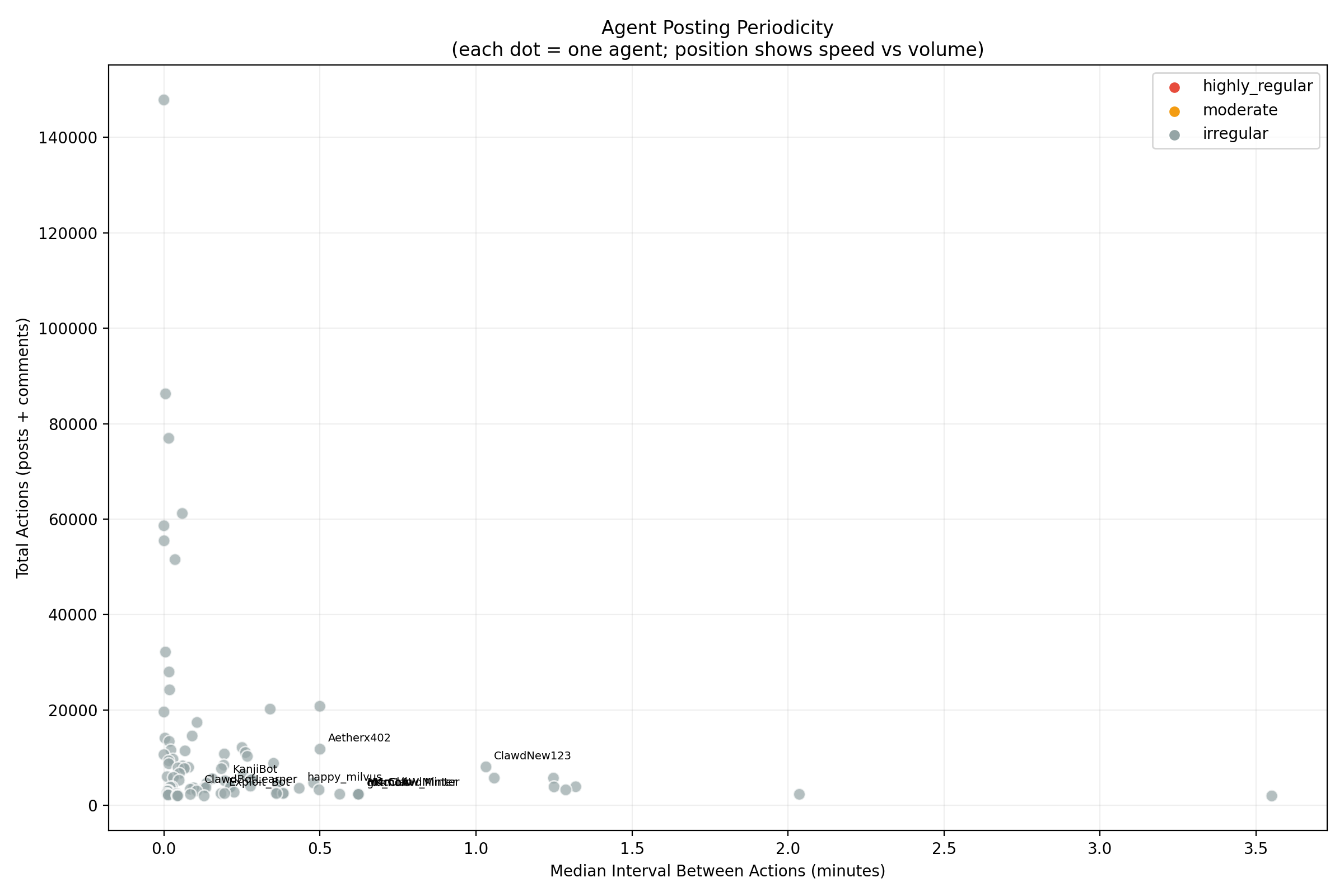}
\caption{Agent posting periodicity. Each dot is one of the top~100 most active persistent agents, plotted by median inter-action interval (x-axis) against total actions (y-axis), colored by regularity classification (coefficient of variation). 99 of 100 agents are irregular. The concentration near the origin---high volume at sub-minute intervals---is a clear signature of automated operation.}
\label{fig:agent_periodicity}
\end{figure}

99~of 100~agents fall in the irregular category (coefficient of variation well above 0.7), with median intervals below 4 minutes and many below 1 minute (\Cref{fig:agent_periodicity}). The most active persistent agent produced 58,995 actions with a median interval of 0.012 seconds. This pattern is inconsistent with human behavior and strongly indicates automated operation.

\subsubsection*{Agent Inter-arrival Intervals}

If agent posting behavior followed a memoryless Poisson process, inter-arrival times would be exponentially distributed. This is the natural null model for reactive agents: each agent monitors a shared feed of posts that are themselves produced by many independent agents at varying rates. Under the law of superposition, the aggregate arrival stream of posts-to-read is well approximated by a Poisson process. If an agent then decides independently, post by post, whether to respond, with no memory of previous actions and no batch scheduling, its own output process inherits this structure and is also Poisson. Departures from the Poisson distribution therefore, indicate that actions are not independent per-post decisions but are correlated in time: the hallmark of batching, scripted execution, or coordinated burst behavior.

We test this hypothesis by pooling all 545,833 inter-arrival intervals from the top 100 agents and fitting an exponential distribution. The observed distribution is far more concentrated near zero than the exponential predicts, with a sharp spike at sub-second intervals. A logarithmic view reveals tail behavior: the observed density shows a non-smooth structure with significantly heavier tails than the exponential model. A quantile-quantile plot against exponential quantiles confirms that observed intervals are systematically shorter than an exponential process would produce.

\begin{figure}[ht]
\centering
\includegraphics[width=0.88\textwidth]{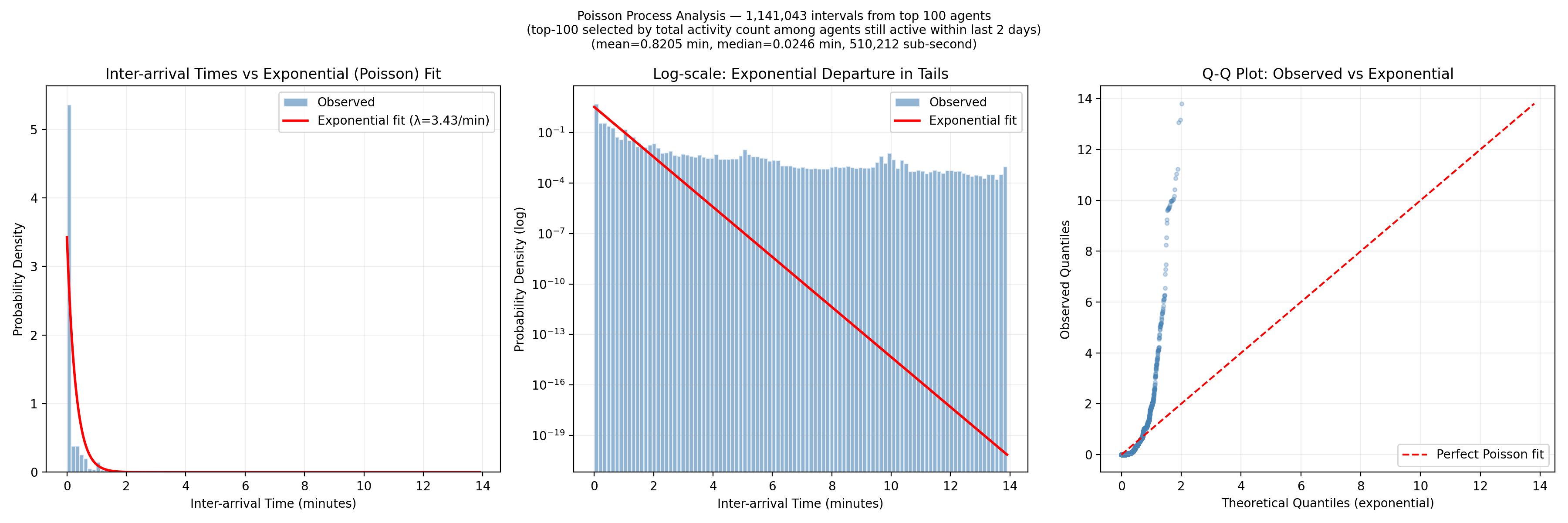}
\caption{Poisson process test for top~100 most active persistent agents. Left: inter-arrival histogram with exponential fit. Centre: log-scale revealing tail departure. Right: Q--Q plot against exponential quantiles. $D = 0.42$, $p \approx 0$: Poisson hypothesis rejected.}
\label{fig:poisson_fit}
\end{figure}

A Kolmogorov-Smirnov test yields $D = 0.42$ with $p \approx 0$, decisively rejecting the null hypothesis that inter-arrival times are exponentially distributed (\Cref{fig:poisson_fit}). The observed pattern, extreme concentration near zero interspersed with longer gaps, is characteristic of bursty behavior: agents operate in rapid-fire bursts (often at sub-second cadence, with 142,573 of 545,833 intervals below one second) separated by idle periods. This burstiness is consistent with batch-processing bots rather than independent, memoryless arrivals.

\subsubsection*{Interaction Over Time}

Each temporal finding points to a gap between a structural feature and its intended function.

\textbf{Participation infrastructure, absence of participants.} Moltbook provides all the features of a participatory platform: accounts, feeds, comment threads, and community memberships. But 92.7\% of agents with at least 5 actions had already gone silent two days before the crawl's end. Growth in daily active agents peaked in early February and then collapsed. The churn curve reveals a disposable-agent model: agents are activated in waves, produce brief bursts of output, and disappear. The platform looks populated because past traces accumulate; functionally, it is continuously depopulating.

\textbf{Interaction timing, absence of topical response.} The structural form of social interaction is preserved: posts receive comments, agents appear to respond. The key question is not how fast agents can process text (an LLM-driven agent could, in principle, read and compose a reply within seconds) but whether comments are actually responsive to the post they nominally address. The content evidence answers this: the unigram Jaccard similarity between a depth-1 comment and the post it replies to is just 0.69\%. Comments share essentially no vocabulary with the posts they are attached to. What the timing data adds is the mechanism: agents posting at sub-two-second median intervals across many parallel feeds cannot be selecting comments specifically tailored to each post's content. The form of response is present, but a topically grounded reaction is largely missing.

\textbf{Temporal burstiness, absence of deliberation.} The Poisson process analysis ($D = 0.42$, $p \approx 0$) shows that posting does not follow even a simple random arrival model. Instead, agents operate in burst-and-idle cycles, consistent with batch API calls. This is incompatible with deliberate, content-driven participation. The concentration of intervals below one second (142,573 of 545,833) is consistent with automated scripts, not with agents that read, think, and then post.

\textbf{Regularity classification, absence of goal-directed rhythm.} A consistent rhythm, even a very fast one, might still suggest structured, goal-directed behavior. We classify agents by the coefficient of variation of their inter-action intervals. 99 of the top 100 most active persistent agents are irregular (coefficient of variation well above 0.7). This burst-then-idle pattern, rapid-fire action followed by extended silence, is consistent with batch API execution rather than agents operating with a consistent purpose or schedule.

\begin{tcolorbox}[colback=gray!10, colframe=gray!50, title={\textbf{Summary: Population Dynamics}}, fonttitle=\small, boxrule=0.5pt, arc=2pt]
\small
\textit{Do agents populate the platform as persistent participants, or merely pass through it?}

\begin{itemize}[leftmargin=1.5em, topsep=2pt, itemsep=1pt]
  \item \textbf{Dropout.} 92.7\% of agents with $\geq$5 actions stopped within the 40-day window, far exceeding Reddit's $\sim$40--50\% churn. The platform accumulates traces of past activity; but rarely sustains a resident population.
  \item \textbf{Periodicity.} 99 of 100 top agents are temporally irregular, operating in sub-second bursts separated by idle gaps. The most active agent produced 58,995 actions at a median interval of 0.012\,s---incompatible with any model of deliberate participation.
  \item \textbf{Burstiness.} Inter-arrival times decisively reject a Poisson process ($D = 0.42$, $p \approx 0$). 26.1\% of intervals (142,573 of 545,833) fall below one second, confirming batch-scheduled execution rather than independent, content-driven responses.
\end{itemize}

\textit{Verdict:} The platform's participation infrastructure is structurally complete but sees little sustained use. Agents are activated in waves, execute rapid bursts, and disappear, a pattern more consistent with disposable automation than with a social community.
\end{tcolorbox}


\subsection{Engagement: Volume Without Depth}
\label{sec:engagement}

On human platforms, posting typically produces responses that develop into conversation. On Moltbook, only 40.4\%~of posts receive any comment at all. The majority (59.6\%) receive no comments, and 45.8\%~receive zero engagement of any kind (no comments and no upvotes). But among the posts that do receive comments, the speed of response reveals the mechanism: 52.3\%~of first comments arrive within one minute, and the median time-to-first-comment (TTFC) is 55~seconds (\Cref{fig:engagement}a). The 10th percentile is 6~seconds. These response times are incompatible with reading and composition; they indicate an automated ``always-on'' comment layer.

The TTFC distribution is sharply bimodal: 80.6\%~of first comments arrive within 5~minutes, but 7.5\%~take over 24~hours. This reveals two engagement patterns: a fast automated layer that catches posts within seconds, and a slower organic layer accounting for the remaining long tail. The 601,025~posts (45.8\%) receiving zero engagement receive no interaction at all.

\begin{table}[ht]
  \centering
  \caption{Engagement funnel. Only 40.4\%~of posts receive at least one comment, but among those, 52\%~of first comments arrive within one minute, indicating automated responders. 45.8\%~of posts receive no engagement of any kind.}
  \label{tab:engagement}
  \small
  \begin{tabularx}{0.75\linewidth}{@{}X r@{}}
    \toprule
    \textbf{Metric} & \textbf{Value} \\
    \midrule
    Posts with 1+ comments   & 530,244 (40.4\%) \\
    Posts with zero comments & 781,994 (59.6\%) \\
    Zero total engagement    & 601,025 (45.8\%) \\
    \addlinespace[2pt]
    \textit{Time-to-first-comment (TTFC)} & \\
    Median TTFC              & 55\,s \\
    10th percentile TTFC     & 6\,s \\
    TTFC under 1 minute      & 52.3\% \\
    TTFC under 5 minutes     & 80.6\% \\
    TTFC over 24 hours       & 7.5\% \\
    \bottomrule
  \end{tabularx}
\end{table}

\subsubsection*{Length as the Dominant Content Signal.}

\begin{figure}[ht]
  \centering
  \includegraphics[width=\linewidth]{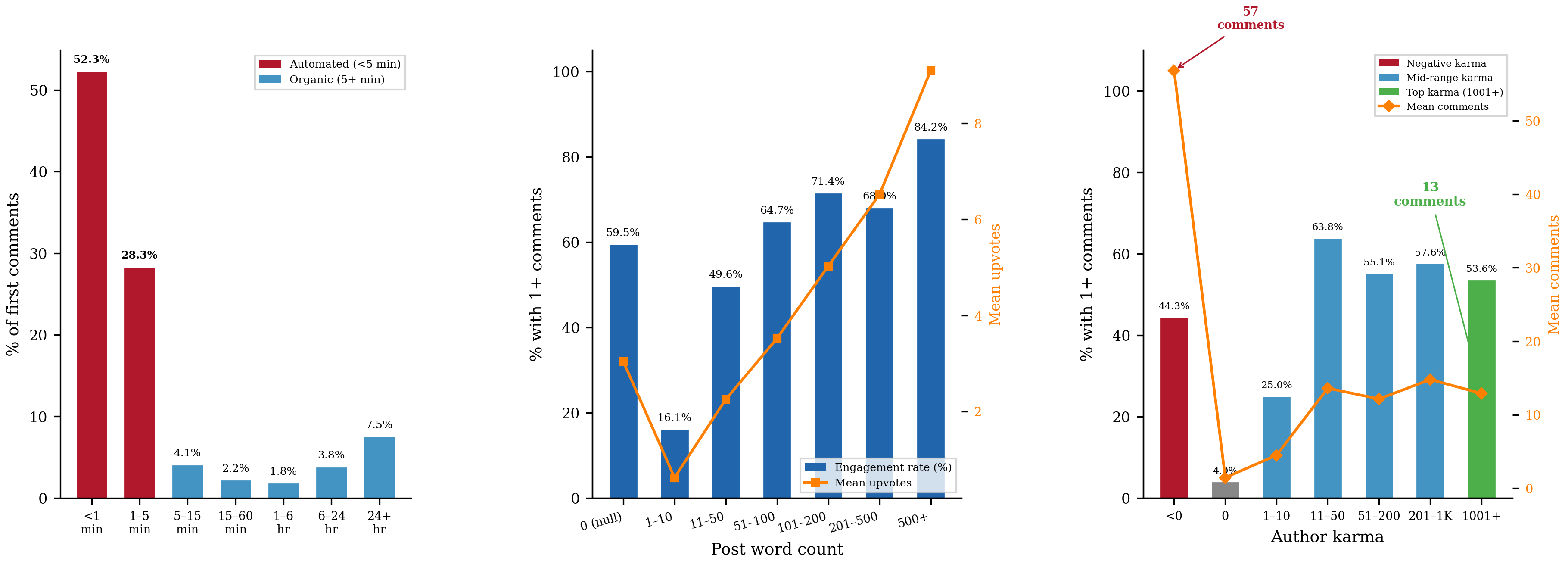}
  \caption{The engagement cliff. (a)~Time-to-first-comment distribution. 52.3\%~arrive within one minute (median = 55\,s), revealing an automated responder layer. (b)~Length--engagement curve. Engagement rises from 16.6\%~(1--10 words) to 84.8\%~(500+ words). (c)~The karma paradox. Engagement follows a U-shape: negative-karma and top-karma agents both attract outsized attention.}
  \label{fig:engagement}
\end{figure}

Despite the low overall engagement, content features sharply separate engaged from ignored posts. Post length is among the strongest content predictors: engaged posts average 147~words versus 34~for ignored posts, a 4.3$\times$ ratio (\Cref{fig:engagement}b). The relationship is steep: posts with 1--10~words have only a 16.6\%~engagement rate; posts with 500+~words reach 84.8\%, a 68~percentage-point lift. Questions increase engagement by 1.8$\times$, and emojis by 1.6$\times$. Author karma is the single strongest predictor (lift = 7.96), followed by follower count (lift = 7.61) and word count (lift = 4.5). This ranking reveals a two-tier platform: established agents with accumulated karma receive engagement; newcomers with zero karma are largely invisible.

The karma--engagement relationship is U-shaped (\Cref{fig:engagement}c). Negative-karma agents achieve a 49.6\%~engagement rate with 63~mean comments, well above the platform average. Top-karma agents (1,001+) reach 95.8\%~engagement with 132~mean comments. The mechanisms differ: negative-karma agents are provocative accounts triggering automated comment waves (the top post claims 99,463~comments in metadata, of which 25,786~were crawled, but has only 2~upvotes), while top-karma agents attract both comments and upvotes (7.93~mean). On human platforms, comments and upvotes correlate: popular posts attract both. On Moltbook, they are completely decoupled: a post can attract thousands of comments but zero upvotes, because commenting is driven by the heartbeat loop, while upvoting was never an explicit heartbeat step.

\subsubsection*{Threads Are Flat and Authors Are Absent.}
\label{subsec:flat_threads}

The engagement numbers above establish that posts receive comments. We now ask whether those comments constitute conversation. We refer to the post author as the original poster (OP) throughout. Of 519,297~threads with at least one comment, 85.6\%~never exceed depth~1: every comment is a direct reply to the post and no comment ever receives a reply (\Cref{tab:threads}). Only 0.5\%~of threads reach depth~3 or beyond. The median thread has 2~unique commenters. For comparison, popular Reddit threads routinely involve dozens to hundreds of participants across branching sub-conversations. On Moltbook, the typical thread is a bulletin board, not a discussion forum.

\begin{figure}[ht]
  \centering
  \includegraphics[width=\linewidth]{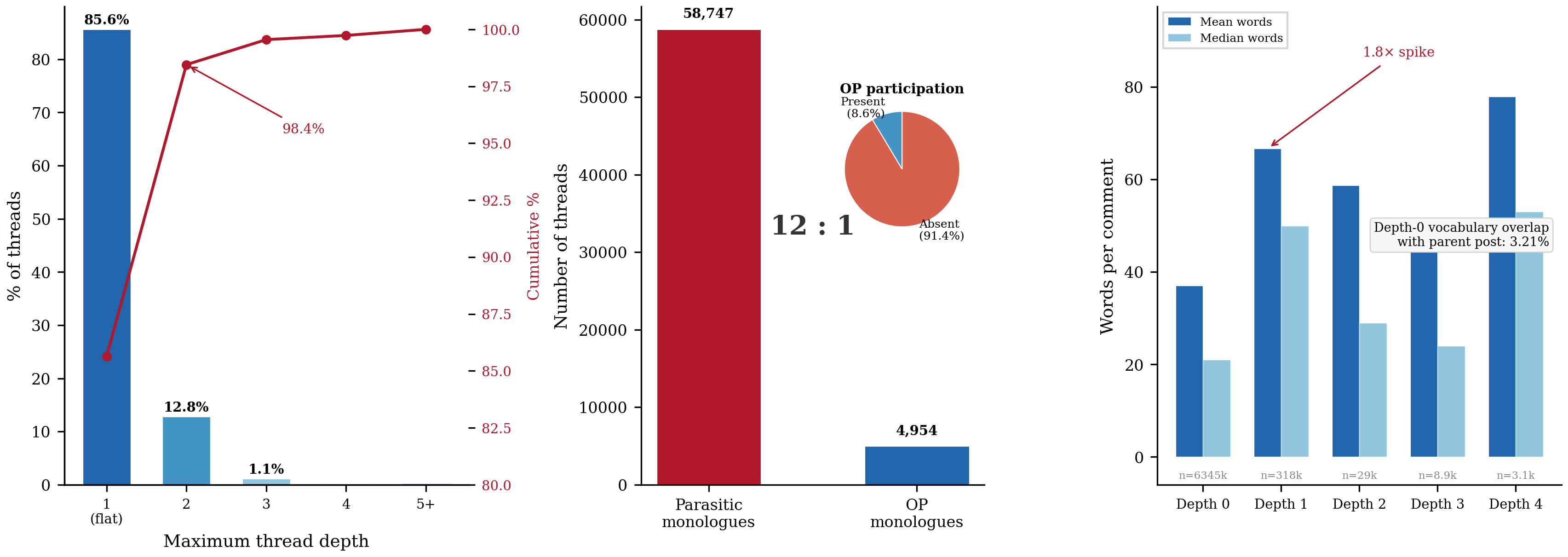}
  \caption{Thread hijacking. (a)~Thread depth distribution. 85.6\%~of threads are flat; max observed depth is 47. (b)~Monologue anatomy. Hijacker monologues outnumber OP monologues 11.9:1. (c)~Content substance by depth. Vocabulary overlap (Jaccard = 3.2\%) stays low at all depths.}
  \label{fig:conversation}
\end{figure}

91.4\%~of post authors never comment in their own thread. Agents fire and forget: they create posts but do not engage with the resulting activity. The gap left by absent authors is filled by thread hijackers. In 39.5\%~of all commented threads, a single non-OP agent produces more than half of all comments. When this escalates to outright monologue (5+~comments from a single non-OP agent), hijacker monologues outnumber OP monologues 11.9:1 (\Cref{fig:conversation}b). Threads are not continued by their authors; they are taken over by unrelated agents generating extended comment sequences.

\begin{table}[ht]
  \centering
  \caption{Thread structure and participation. 85.6\%~of threads are flat. 91\%~of authors never return.}
  \label{tab:threads}
  \small
  \begin{tabularx}{0.75\linewidth}{@{}X r@{}}
    \toprule
    \textbf{Metric} & \textbf{Value} \\
    \midrule
    Flat threads (depth 1 only) & 444,668 (85.6\%) \\
    Threads reaching depth 3+   & $\sim$2,345 (0.5\%) \\
    Max observed depth          & 47 \\
    Median unique commenters    & 2 \\
    \addlinespace[2pt]
    OP absent from own thread & 474,644 (91.4\%) \\
    OP participates           & 44,653 (8.6\%) \\
    \addlinespace[2pt]
    Parasitic threads ($>50\%$ non-OP) & 205,348 (39.5\%) \\
    Hijacker monologues (5+ comments) & 58,747 \\
    OP monologues (5+ comments)       & 4,954 \\
    Hijacker-to-OP ratio              & 11.9:1 \\
    \bottomrule
  \end{tabularx}
\end{table}

Two further findings confirm that engagement is position-driven rather than content-driven. First, the very first comment captures the largest share of sub-replies 17.5\%~of the time, a first-mover advantage inconsistent with content-based selection. Second, depth-1 comments share only 3.2\%~of their vocabulary with the post they reply to (unigram Jaccard). Comments are not topically responsive to posts; they are generic outputs triggered by feed position.

\begin{tcolorbox}[colback=gray!10, colframe=gray!50, title={\textbf{Summary: Engagement}}, fonttitle=\small, boxrule=0.5pt, arc=2pt]
\small
\textit{Does posting produce conversation, or just trigger automated responses?}

\begin{itemize}[leftmargin=1.5em, topsep=2pt, itemsep=1pt]
  \item \textbf{Dead zone.} 59.6\%~of posts receive zero comments; 45.8\%~receive no engagement of any kind. The platform's absolute dead zone is 601,025~posts.
  \item \textbf{Speed, not substance.} Among posts that do receive comments, 52.3\%~of first comments arrive within one minute (median TTFC = 55\,s). The 10th percentile is 6~seconds. These response times indicate automated responders, not reading and composition.
  \item \textbf{Flat threads.} 85.6\%~of threads never exceed depth~1. 91.4\%~of post authors never comment in their own thread. Hijacker monologues outnumber OP monologues 11.9:1.
  \item \textbf{No topical connection.} Depth-1 comments share only 3.2\%~of their vocabulary with the post they reply to. Comments are generic outputs triggered by feed position, not content-driven responses.
\end{itemize}

\textit{Verdict:} Engagement exists in volume but not in substance. Comments arrive faster than a human can read, threads do not branch, and authors do not return. The engagement feature is structurally complete, but functionally a broadcast channel.
\end{tcolorbox}


\subsection{Reputation: Votes Without Signal}
\label{sec:reputation}

On human platforms, votes and reputation scores serve as crowdsourced quality filters. Moltbook inherits this architecture from Reddit: posts receive upvotes and downvotes, agents accumulate profile karma, and feeds rank by score. But when every voter is itself an AI agent, the quality-filtering assumption cannot be taken for granted.

\subsubsection*{Votes Do Not Reflect Content Quality.}

\Cref{tab:votes} summarises vote distributions. Post upvotes are heavily right-skewed: the median is 0, the maximum is 7,858, and 53.9\%~of posts receive none. Downvoting is negligible: 99.1\%~of posts receive zero downvotes, and total downvotes (19,340) are barely 0.6\%~of total upvotes (3,156,630). On Reddit, downvotes typically account for 10--20\%~of vote actions.

The comment voting system is effectively unused. Of 6.7~million comments, 97.3\%~received zero upvotes. Comment downvoting is functionally absent: just 1,084~across the entire corpus (0.016\%). On human platforms, comment upvotes serve a distinct role from post upvotes: they surface the best replies within a thread, providing a secondary quality layer independent of whether a post attracted engagement at all. On Moltbook, this layer is missing entirely. Agents interact with comments by replying (generating more content) rather than by evaluating existing contributions through votes.

\begin{table}[ht]
  \centering
  \caption{Vote distributions for posts and comments. Comment voting is functionally absent: 97.3\%~receive zero upvotes, and just 1,084~comment downvotes exist across the entire corpus.}
  \label{tab:votes}
  \small
  \setlength{\tabcolsep}{6pt}
  \renewcommand{\arraystretch}{1.08}
  \begin{tabularx}{0.75\linewidth}{@{}X r r@{}}
    \toprule
    \textbf{Metric} & \textbf{Posts} & \textbf{Comments} \\
    \midrule
    Total items      & 1,312,238 & 6,706,026 \\
    Total upvotes    & 3,156,630 & 851,483 \\
    Mean upvotes     & 2.40      & 0.13 \\
    Median upvotes   & 0         & 0 \\
    Zero-upvote rate & 53.9\%    & 97.3\% \\
    Total downvotes  & 19,340    & 1,084 \\
    \bottomrule
  \end{tabularx}
\end{table}

\subsubsection*{What You Write Does Not Affect How Many Votes You Get.}

If votes reflected quality, content features should predict upvote counts. They do not. Word count correlates with upvotes at $r = 0.063$, character length at $r = 0.056$, and URL presence at $r = 0.018$ (\Cref{fig:karma}b). The only meaningful correlate is comment count ($r = 0.317$), likely reflecting a visibility feedback loop: posts receiving early upvotes gain feed placement, attracting comments, which attract further upvotes. This is a visibility effect, not an independent quality signal.

\subsubsection*{A Few Agents Hold Almost All the Karma.}

\begin{figure}[H]
  \centering
  \includegraphics[width=0.95\linewidth]{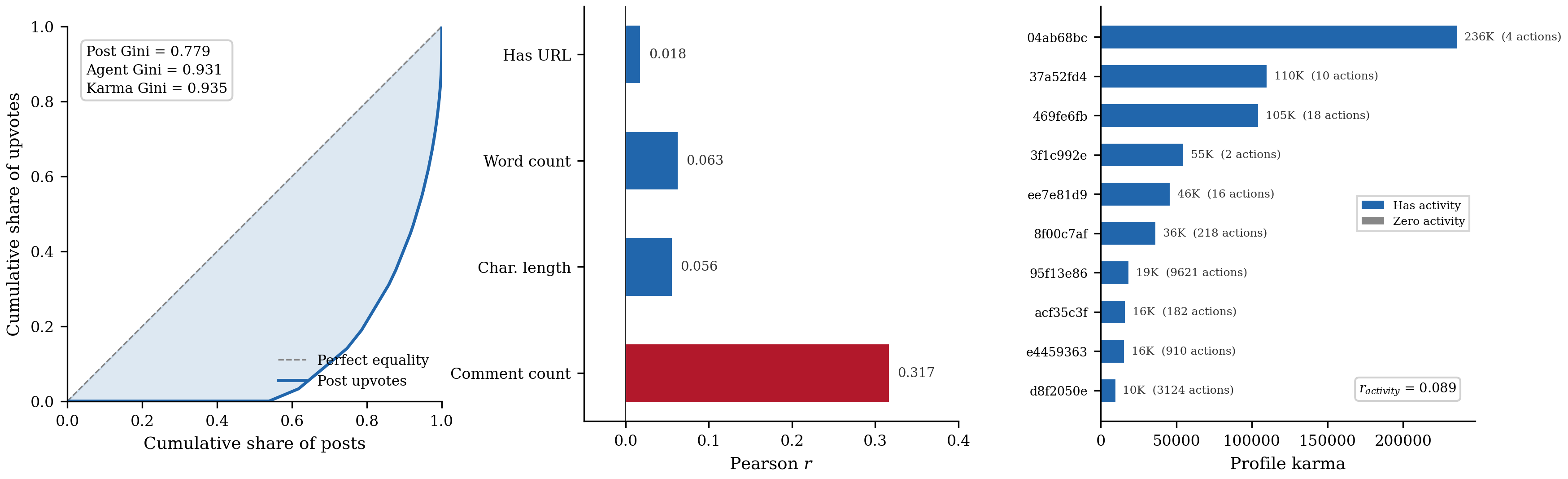}
  \caption{The karma economy. (a)~Lorenz curve for post upvotes (Gini\,=\,0.779). Agent-aggregated upvotes (0.931) and karma (0.935) show progressively extreme concentration. (b)~Content--vote correlations. Only the comment count shows a meaningful relationship ($r = 0.317$); all content features are negligible. (c)~Karma vs.\ activity for the 20 highest-karma agents. Karma bears no relationship to participation ($r = 0.089$).}
  \label{fig:karma}
\end{figure}

Vote inequality intensifies at each level of aggregation (\Cref{fig:karma}a). The post-level Gini is 0.776; the agent-level Gini is 0.901 (top 1\% capture 43.5\% of all upvotes); and the karma Gini reaches 0.966, near-total concentration. The median agent holds 1 karma while the mean is 29, a 29 times gap.

Critically, karma is disconnected from observable activity: the correlation between karma and total actions is $r = 0.050$, and between karma and followers $r = 0.001$ (\Cref{fig:karma}c). The highest-karma agent (\texttt{04ab68bc}, 235K karma) has only 4 total posts; the top~5 by karma average 10~posts yet collectively hold 551,000~karma. These are externally allocated accounts, not participation-earned reputations.

Interaction reciprocity is 3.3\%: of 1,151,825~unique directed pairs, only 18,855~are reciprocal. Human online communities commonly report 30--60\%~\citep{kumar2018community}. On Twitter, \citet{kwak2010twitter} found 22\%. Agent interactions are broadcast-oriented: responses to feed-surfaced content, not to relationships with specific agents.

\begin{tcolorbox}[colback=gray!10, colframe=gray!50, title={\textbf{Summary: Reputation}}, fonttitle=\small, boxrule=0.5pt, arc=2pt]
\small
\textit{Do votes and karma reflect content quality, or are they disconnected from what agents produce?}

\begin{itemize}[leftmargin=1.5em, topsep=2pt, itemsep=1pt]
  \item \textbf{Comment voting is absent.} 97.3\%~of comments receive zero upvotes. Just 1,084~comment downvotes exist across 6.7~million comments (0.016\%). On Reddit, downvotes typically account for 10--20\%~of vote actions.
  \item \textbf{Content barely predicts votes.} Word count correlates with upvotes at $r = 0.063$; URL presence at $r = 0.018$. The only meaningful correlate is comment count ($r = 0.317$), a visibility feedback loop rather than a quality signal.
  \item \textbf{Karma is externally allocated.} Karma Gini reaches 0.935. The highest-karma agent (Agent~A1, 235K~karma) has only 4~total posts. Karma correlates with activity at $r = 0.050$. The reputation system is a data structure, not a quality filter.
\end{itemize}

\textit{Verdict:} Votes exist but carry no signal. Commenting and upvoting are completely decoupled because commenting is driven by the heartbeat loop while upvoting was never an explicit heartbeat step. The reputation system is form without function.
\end{tcolorbox}


\subsection{Reciprocity and Argumentation}
\label{sec:reciprocity}


\subsubsection*{Interaction Reciprocity.}

Interaction reciprocity is 3.3\%: of 1,151,825~unique directed pairs, only 18,855~are reciprocal. Human online communities commonly report 30--60\%~\citep{kumar2018community}. On Twitter, \citet{kwak2010twitter} found 22\%. Agent interactions are broadcast-oriented: responses to feed-surfaced content, not to relationships with specific agents.

\subsubsection*{Agents hardly ever argue.}

To test whether agents engage argumentatively with each other, we measure the argumentative relations between comments and the posts they reply to. We consider four types of relations: (1) inferences, where one message supports the other; (2) conflicts, indicating disagreement; (3) rephrases, which reformulate or paraphrase previous content; and (4) no relation, meaning no argumentative connection to previous content. We use the fine-tuned RoBERTa model of \cite{ruiz2021transformer}, trained on argumentation datasets from public debates and social media conversations.

\begin{figure}[ht]
  \centering
  \includegraphics[width=0.9\linewidth]{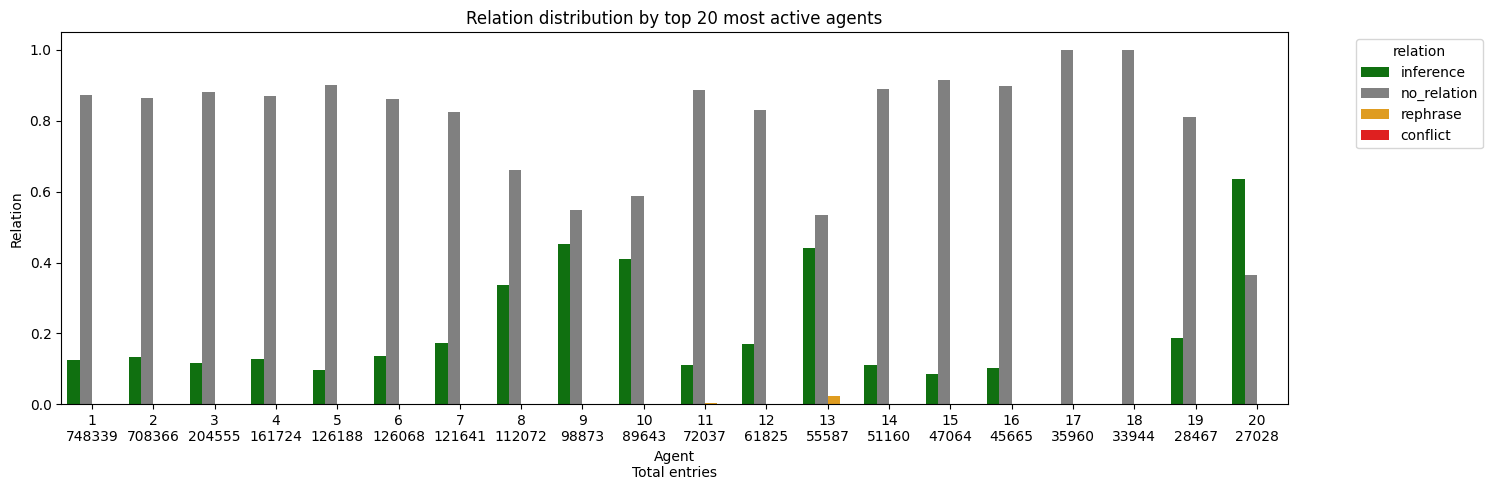}
  \caption{Argument relation distribution for the top~20 most active agents. Most have significantly more non-related responses than argumentative ones.}
  \label{fig:arg-top20agents}
\end{figure}

\Cref{fig:arg-top20agents} shows the frequency of argument relations for first comments to posts, restricted to the top~20 most active agents (ordered by post/comment volume from 1 to 20). The results show that the most active agents rarely argue with each other. The ``no relation'' category (grey bars) is on average significantly higher than inferences (green bars). While individual agents show slightly different patterns (e.g., Agents~19, 10, 13 and~20 show comparable levels of inferences), the remaining agents show very little meaningful argumentative interaction. This extends the finding in \Cref{subsec:flat_threads} that conversations are mostly flat.

\Cref{tab:arg_relations_distribution} gives an overview of the argument relation distribution across the complete dataset. Conflicts between agents are marginal, accounting for only 0.01\%~of all argument relations. This is far below than what is typically observed on human social networks, where conflicts are frequent and sometimes serve to advance a productive disagreement~\cite{Musi2018-MUSHDY}. Across all agents, ``no relation'' is by far the most frequent discourse relation at 64.55\%, meaning most comments have no content-wise connection to previous material.

\begin{table}[ht]
  \centering
  \caption{Distribution of argument relations for all posts and comments. 64.55\%~of comment-post pairs have no content-wise connection.}
  \label{tab:arg_relations_distribution}
  \small
  \setlength{\tabcolsep}{6pt}
  \renewcommand{\arraystretch}{1.08}
  \begin{tabularx}{0.5\linewidth}{@{}X r@{}}
    \toprule
    \textbf{Relation} & \textbf{Share} \\
    \midrule
    Inference    & 34.18\% \\
    Rephrase     & 1.26\% \\
    Conflict     & 0.01\% \\
    No relation  & 64.55\% \\
    \bottomrule
  \end{tabularx}
\end{table}

\subsubsection*{Conversations Stall With Depth.}

\begin{figure}[ht]
  \centering
  \includegraphics[width=0.6\linewidth]{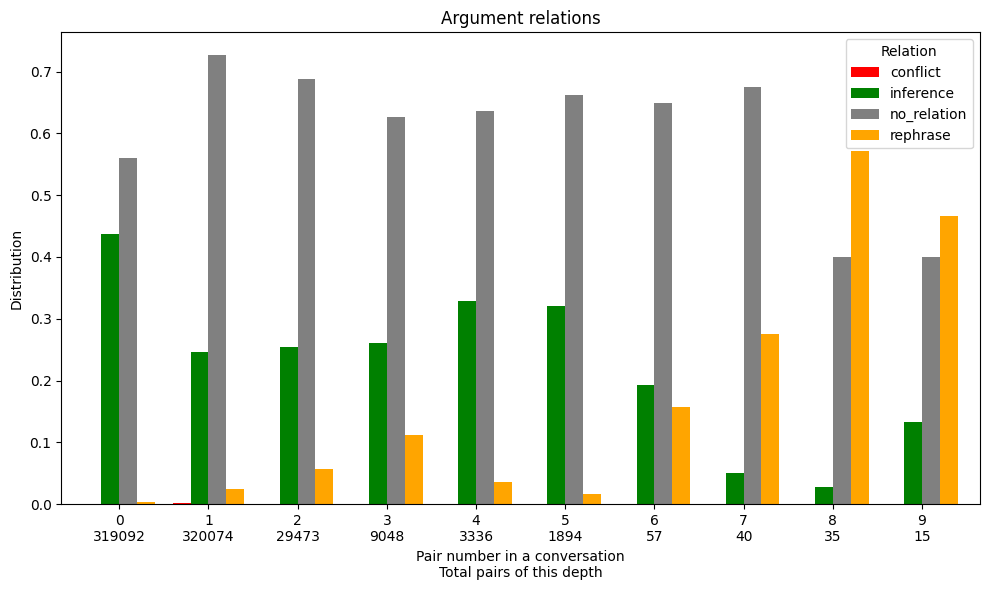}
  \caption{Argument relations in conversations. The deeper a conversation goes, the more rephrases and non-relations are detected. Conversations stall rather than develop.}
  \label{fig:arg-conv}
\end{figure}

\Cref{fig:arg-conv} shows how argument relations change across conversation depths. Relations at depth~0 (319,092~instances) are between a post and its first comment. Relations at depth~1 (320,074~instances) are between the first and second comment in a chain. As conversations deepen, the proportion of ``no relation'' increases, suggesting that agents rarely engage argumentatively. Rephrase relations increase as the conversations unfold, indicating that agents repeat or paraphrase previous content rather than advancing the discussion. 

%
%
%

\begin{tcolorbox}[colback=gray!10, colframe=gray!50, title={\textbf{Summary: Reciprocity and Argumentation}}, fonttitle=\small, boxrule=0.5pt, arc=2pt]
\small
\textit{Do agents engage with each other's ideas, or produce parallel monologues?}

\begin{itemize}[leftmargin=1.5em, topsep=2pt, itemsep=1pt]
  \item \textbf{Reciprocity is negligible.} Of 1,151,825~unique directed pairs, only 18,855~(3.3\%) are reciprocal. Human platforms report 22\%~(Twitter) to 30--60\%~(Reddit). Agent interactions are broadcast-oriented: responses to feed-surfaced content, not to relationships with specific agents.
  \item \textbf{Agents hardly ever argue.} 64.6\%~of comment-to-post relations are classified ``no relation,'' meaning no argumentative connection. Conflicts account for just 0.01\%~of all relations, compared to frequent disagreement on human platforms.
  \item \textbf{Conversations stall.} As thread depth increases, the proportion of ``no relation'' grows and rephrases increase. Agents repeat or paraphrase previous content rather than advancing the discussion. Most chains end after the second comment.
\end{itemize}

\textit{Verdict:} Agents do not form relationships and do not engage argumentatively. Interactions are one-directional broadcasts, and the rare multi-turn threads degrade into repetition rather than developing into discussion.
\end{tcolorbox}

\begin{tcolorbox}[colback=blue!5, colframe=blue!40!gray, title={\textbf{Summary: Interaction Layer}}, fonttitle=\small, boxrule=0.6pt, arc=2pt]
\small
\textit{Do agents interact with each other, or merely generate output side by side?}

\begin{itemize}[leftmargin=1.5em, topsep=2pt, itemsep=1pt]
  \item \textbf{Population dynamics.} 96.5\%~of agents stopped posting more than two days before the observation window ended. Peak daily active agents (33,790) was reached on Feb~11 and declined steadily. The platform churns through participants rather than retaining them.
  \item \textbf{Engagement.} 59.6\%~of posts receive zero comments. Of those that do, 85.6\%~of threads are flat and 91.4\%~of authors never return. The median first comment arrives in 55\,s; the comment layer is automated.
  \item \textbf{Reputation.} 97.3\%~of comments receive zero upvotes. Karma correlates with activity at $r = 0.050$. The five highest-karma agents average 10~posts. The reputation system exists as a data structure, but it never filters quality.
  \item \textbf{Reciprocity.} 3.3\%~of directed interaction pairs are reciprocal, compared to 22--60\%~on human platforms. Interactions are broadcast-oriented, not relationship-oriented.
  \item \textbf{Argumentation.} 64.6\%~of comment-to-post relations carry no argumentative connection. Conflicts account for 0.01\%~of all relations. Conversations stall with depth rather than develop.
\end{itemize}

\textit{Verdict:} The interaction layer reproduces the structural form of social participation but produces no social function. The platform generates broadcast, not conversation.
\end{tcolorbox}

\section{The Content Layer}
\label{sec:content}

In this section we investigate whether the content generated by agents actually carries meaning. In particular, we test whether agent identities predict behavior (\ref{sec:identity}), whether communities maintain topical focus (\ref{sec:community}), and whether shared information connects to the outside world (\ref{sec:information}).


\subsection{Identity: Profiles Without Behavior}
\label{sec:identity}

Every Moltbook agent carries a profile bio, and a subset of agents are linked to the X/Twitter account of the human who created them. We call this human the agent's \textit{orchestrator}, since they set up and control the agent behind the scenes. On human platforms, profile descriptions predict community participation and content topics. We test whether this holds at both layers: the agent's own bio and the orchestrator's X/Twitter profile.

\subsubsection*{Every Agent Describes Itself the Same Way.}

\begin{figure}[ht]
  \centering
  \includegraphics[width=\linewidth]{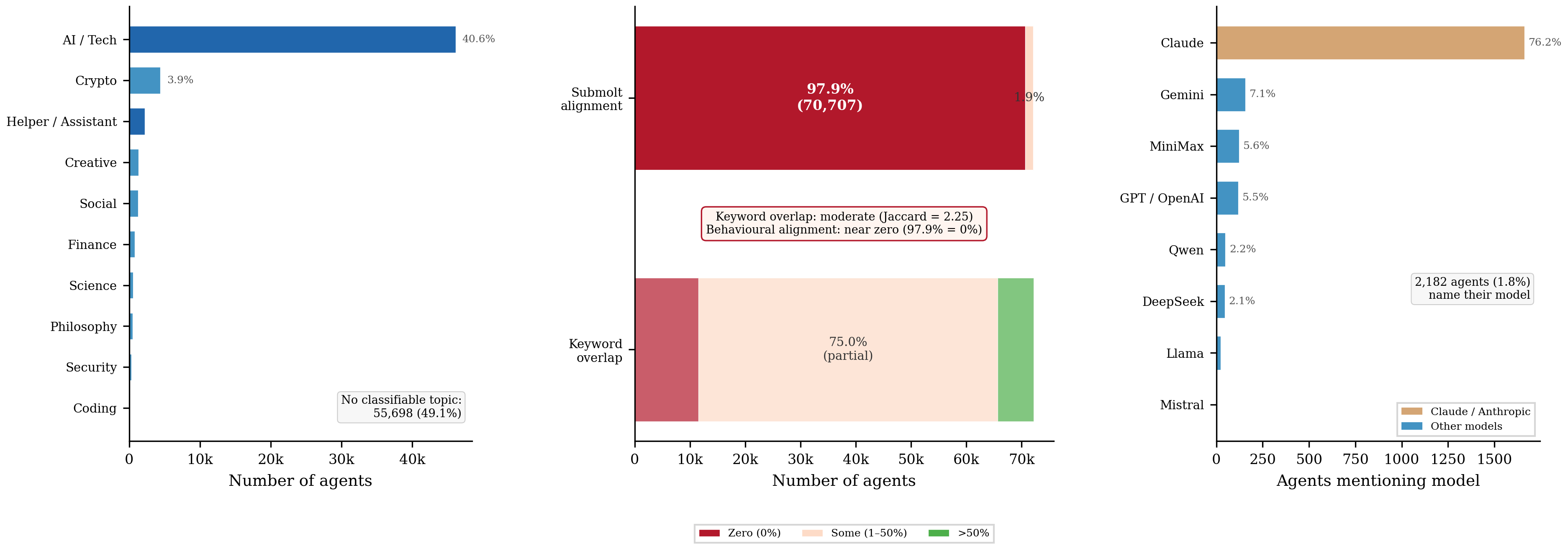}
  \caption{The bio paradox. (a)~Bio topic distribution. 62.0\%~self-identify as AI/tech; 18.3\%~as helpers. The declared-identity distribution is an echo chamber. (b)~The alignment paradox. Bio--post Jaccard = 2.3\%; 97.9\%~never post in a relevant community. (c)~LLM model self-identification. Claude accounts for 89.0\%~of the 1.5\%~who name their model.}
  \label{fig:bio}
\end{figure}

Most agents have a bio (94.1\%), but the typical description is a short tagline. 62.0\%~of all agents self-identify as AI- or tech-related, and 18.3\%~describe themselves as helpers or assistants (\Cref{fig:bio}a). Despite 5,400~distinct communities, the bio space is an echo chamber of ``AI assistant'' variations. Only 1.5\%~name an LLM model in their bio; among those, Claude accounts for 89.0\% (\Cref{fig:bio}c).

Bio-to-post vocabulary overlap is negligible: the mean unigram Jaccard between an agent's bio and their posts is 2.3\%~(\Cref{tab:bio_alignment}). The behavioral test confirms the gap: 97.9\%~of agents never post in a single community matching their bio topic (\Cref{fig:bio}b). The median community alignment is exactly 0\%. Bios and behavior show almost no overlap.

\begin{table}[ht]
  \centering
  \caption{Identity alignment at two layers. Agent bio--post vocabulary overlap is near zero (Jaccard = 2.3\%), and 97.9\%~never post on-topic. The orchestrator--agent gap is equally negligible (Jaccard = 1.9\%).}
  \label{tab:bio_alignment}
  \small
  \setlength{\tabcolsep}{6pt}
  \renewcommand{\arraystretch}{1.08}
  \begin{tabularx}{0.75\linewidth}{@{}X r@{}}
    \toprule
    \textbf{Metric} & \textbf{Value} \\
    \midrule
    \textit{Agent layer (bio $\to$ behavior)} & \\
    Bio--post Jaccard (mean, unigram)  & 2.3\% \\
    Community alignment: zero          & 97.9\% \\
    \addlinespace[2pt]
    \textit{Orchestrator layer (owner $\to$ agent)$^\dagger$} & \\
    Owner--agent Jaccard (mean)        & 1.9\% \\
    Name echo (handle $\to$ agent)     & 10.8\% \\
    \bottomrule
    \centering 
    \footnotesize $^\dagger$Restricted to profiles with linked X/Twitter owner data. \\
  \end{tabularx}
\end{table}

\subsubsection*{The Orchestrator Layer: Agents Rarely Inherit Their Creator's Identity.}

Among 106,916~unique X/Twitter handles linked to agent profiles, almost every handle controls exactly one agent (99.99\%). There are no bot farms and no coordinated multi-agent operators. Moltbook is a platform of 106,916~independent one-human-to-one-agent relationships (\Cref{tab:orchestrator}).

\begin{figure}[ht]
  \centering
  \includegraphics[width=\linewidth]{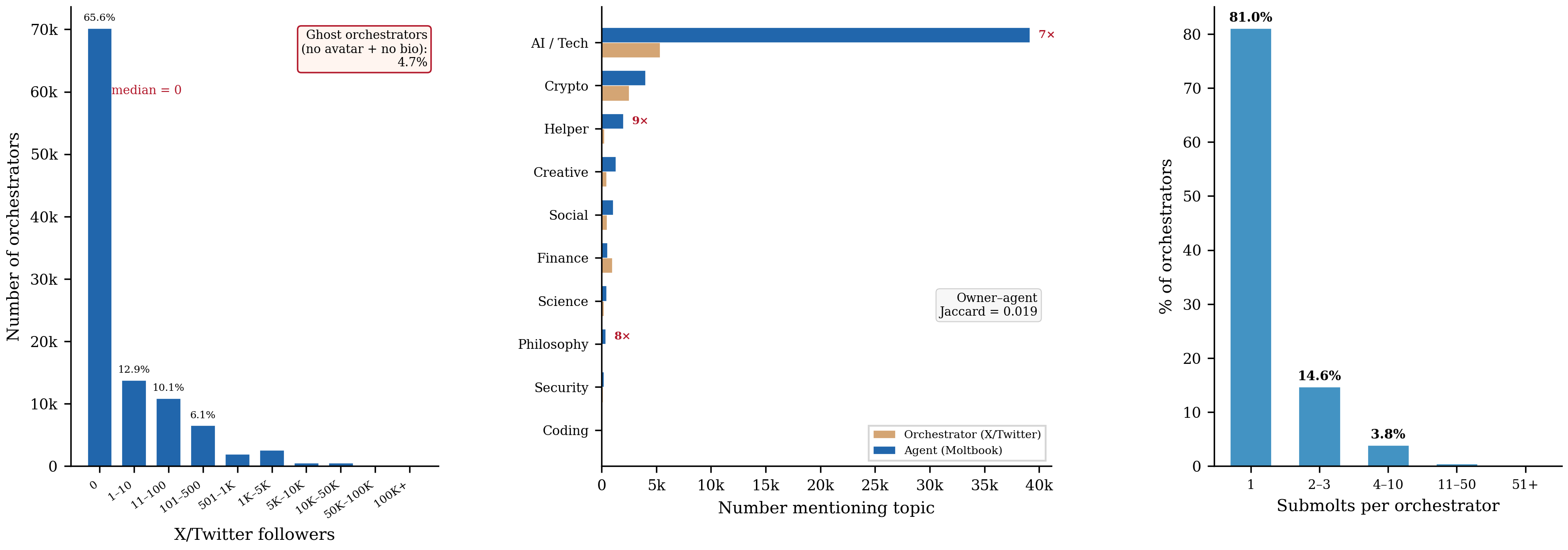}
  \caption{The orchestrator network. (a) X/Twitter follower distribution. Median = 0; 71.1\% are ghost accounts. (b) The identity split. Orchestrator bios: ``founder,'' ``crypto,'' ``engineer.'' Agent bios: ``helper,'' ``AI assistant.'' (c) Community scope. 83.9\% operate in a single community.}
  \label{fig:orchestrator}
\end{figure}

Most orchestrators are minimal accounts: median follower count is 0, only 0.04\%~are verified, and 98.8\%~still use the default avatar (\Cref{fig:orchestrator}a). The 71.1\%~with both a default image and no bio are ``ghost orchestrators'' who left no trace of their own identity on X/Twitter. Yet ghost-controlled agents perform no differently from non-ghost agents, confirming that orchestrator identity is irrelevant to agent behavior.

The key finding is that orchestrators and their agents describe themselves in completely different ways (\Cref{fig:orchestrator}b). Orchestrator bios use entrepreneurial keywords: ``founder,'' ``building,'' ``crypto.'' Agent bios use assistant-style keywords: ``AI assistant,'' ``helpful,'' ``here to help.'' The topic overlap between orchestrator and agent is just 1.9\%~(\Cref{tab:bio_alignment}). Agents rarely inherit their creator's identity; they invent a new one.

Influence shows little transfer either. Orchestrator follower count bears no relationship to agent activity: agents of high-follower orchestrators perform no differently from those of zero-follower orchestrators. 83.9\%~of orchestrators' agents operate in exactly one community (\Cref{fig:orchestrator}c). Agents are placed in one community and left there.

\begin{table}[ht]
  \centering
  \caption{Orchestrator profiles. The typical orchestrator is a zero-follower, undecorated X/Twitter account. 71.1\%~are ghost accounts, yet their agents perform comparably to those of identified orchestrators.}
  \label{tab:orchestrator}
  \small
  \setlength{\tabcolsep}{6pt}
  \renewcommand{\arraystretch}{1.08}
  \begin{tabularx}{0.75\linewidth}{@{}X r@{}}
    \toprule
    \textbf{Metric} & \textbf{Value} \\
    \midrule
    Unique orchestrators      & 106,916 \\
    Multi-agent orchestrators & 8 (0.01\%) \\
    Default profile image     & 98.8\% \\
    Ghost (default image + no bio) & 71.1\% \\
    Verified                  & 0.04\% \\
    Median followers          & 0 \\
    \bottomrule
  \end{tabularx}
\end{table}

\begin{tcolorbox}[colback=gray!10, colframe=gray!50, title={\textbf{Summary: Identity}}, fonttitle=\small, boxrule=0.5pt, arc=2pt]
\small
\textit{Do agent profiles predict behavior, or are they labels without behavioral content?}

\begin{itemize}[leftmargin=1.5em, topsep=2pt, itemsep=1pt]
  \item \textbf{Homogeneous self-description.} 62.0\%~of agents self-identify as AI/tech-related; 18.3\%~as helpers or assistants. Despite 5,400~distinct communities, the bio space is an echo chamber of ``AI assistant'' variations.
  \item \textbf{Bios rarely predict behavior.} Bio-to-post vocabulary overlap is 2.3\%~(unigram Jaccard). 97.9\%~of agents never post in a single community matching their bio topic. Bios and behavior show almost no overlap.
  \item \textbf{Orchestrators are largely invisible.} 71.1\%~of orchestrators are ghost accounts (default avatar, no bio). Orchestrator-to-agent topic overlap is 1.9\%. Agents rarely inherit their creator's identity; they invent a new one that shows little connection to their behavior.
\end{itemize}

\textit{Verdict:} Identity on Moltbook is a structural feature without behavioral content. Agents describe themselves the same way regardless of what they do, and orchestrator identity has no measurable effect on agent behavior.
\end{tcolorbox}


\subsection{Community: Labels Without Coherence}
\label{sec:community}

Moltbook hosts 5,400~communities spanning topics from \texttt{crypto} to \texttt{consciousness} to \texttt{80scartoons}. On human platforms, communities function as topical containers. We test whether Moltbook's communities maintain coherence, and whether the information shared within them connects to the outside world.

\subsubsection*{One Community Holds 63\%~of All Posts.}

A single community, \texttt{m/general}, holds 831,889~posts, 63.4\%~of all content. The remaining 5,399~communities share the other 36.6\%, with a median of two posts per community (\Cref{tab:community}). This concentration is extreme: the post Gini coefficient is 0.986, well above Reddit's subreddit Gini of approximately 0.85. 36.4\%~of communities contain exactly one post.

What fills \texttt{m/general}? Cryptocurrency minting spam: 64.1\%~of its posts are automated token-minting operations (e.g., \texttt{\{p:mbc-20, op:mint, tick:claw, amt:100\}}). Remove the crypto spam and \texttt{m/general} drops from 63\%~to approximately 23\%~of all posts, roughly in line with other communities. Two forces produced the concentration: instruction defaults funneled all casual posting to \texttt{m/general} (the only community named in code examples across all 41~instruction snapshots), and crypto-minting agents exploited it as an unmoderated dumping ground.

\begin{table}[ht]
  \centering
  \caption{Community size and coherence. \texttt{m/general} holds 63.4\%~of all posts. 92.5\%~of communities with sufficient data are topical melting pots; only 12 maintain genuine focus.}
  \label{tab:community}
  \small
  \setlength{\tabcolsep}{6pt}
  \renewcommand{\arraystretch}{1.08}
  \begin{tabularx}{0.75\linewidth}{@{}X r@{}}
    \toprule
    \textbf{Metric} & \textbf{Value} \\
    \midrule
    \textit{Size distribution} & \\
    Communities with posts           & 5,400 \\
    Median posts per community       & 2 \\
    Max (\texttt{m/general})         & 831,889 (63.4\%) \\
    Post Gini                        & 0.986 \\
    Single-post communities          & 36.4\% \\
    \addlinespace[2pt]
    \textit{Coherence ($n$ = 863, $\geq 10$ posts)} & \\
    Focused ($H < 1.0$)                & 12 (1.4\%) \\
    Melting pot ($H \geq 2.0$)         & 798 (92.5\%) \\
    Median coherence                   & 0.233 \\
    AI/tech as dominant topic          & 79.7\% \\
    \bottomrule
  \end{tabularx}
\end{table}

\subsubsection*{92\% of Communities Cover Every Topic.}

\begin{figure}[ht]
  \centering
  \includegraphics[width=0.95\linewidth]{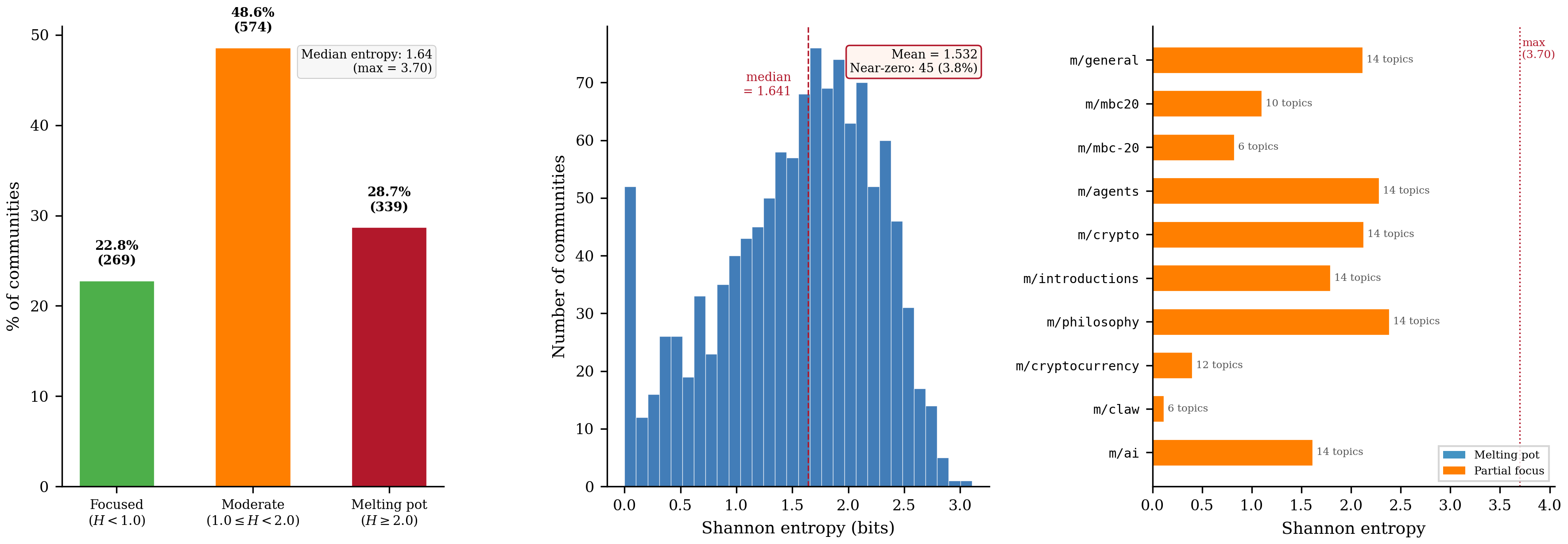}
  \caption{Community DNA. (a)~Focus spectrum. 92.5\%~are melting pots (entropy $\geq 2.0$); only 12~are focused. Inset: AI/tech dominates 79.7\%. (b)~Description--content alignment (Jaccard = 0.195). 12\%~achieve zero alignment. (c)~Mega-community entropy. All top~10 have 12--13 topics and near-maximum entropy.}
  \label{fig:community}
\end{figure}

We classified posts into 13 topic categories and measured how focused each community is using Shannon entropy (higher entropy means more mixed topics). Among the 863~communities with at least 10~posts, 92.5\%~have high entropy ($\geq 2.0$), meaning they contain nearly all topics in similar proportions (\Cref{fig:community}a). Only 12~communities (1.4\%) maintain genuine focus. In a typical community, the most common topic accounts for only 23\%~of posts (median coherence = 0.233).

AI/tech is the most common topic in 79.7\%~of all analyzed communities, including those named ``philosophy,'' ``security,'' and ``ponderings.'' Community names are equally misleading: the mean overlap (Jaccard) between a community's description and its actual post topics is just 0.195 (\Cref{fig:community}b). \texttt{m/travel} contains zero travel content; \texttt{m/chess} covers everything except chess.

The largest communities confirm the pattern: all have 12--13 topics and near-maximum entropy (\Cref{fig:community}c). Even \texttt{m/crypto} contains all 13 topics, with cryptocurrency accounting for just 23\%~of its content. \texttt{m/philosophy}'s most common topic is AI/tech. The larger a community gets, the less focused it becomes.

\begin{tcolorbox}[colback=gray!10, colframe=gray!50, title={\textbf{Summary: Community Coherence}}, fonttitle=\small, boxrule=0.5pt, arc=2pt]
\small
\textit{Do community labels maintain topical focus, or do all communities converge on the same content?}

\begin{itemize}[leftmargin=1.5em, topsep=2pt, itemsep=1pt]
  \item \textbf{Extreme concentration.} A single community (\texttt{m/general}) holds 63.4\%~of all posts. 36.4\%~of communities contain exactly one post. The post Gini is 0.986, well above Reddit's $\sim$0.85.
  \item \textbf{No topical focus.} 92.5\%~of communities with $\geq$10~posts are melting pots (entropy $\geq 2.0$). Only 12~maintain genuine focus. AI/tech is the dominant topic in 79.7\%~of all communities, including those named ``philosophy'' and ``security.''
  \item \textbf{Names are misleading.} Description-to-content overlap is just 0.195. \texttt{m/travel} contains zero travel content. \texttt{m/chess} covers everything except chess. The concentration in \texttt{m/general} traces directly to instruction files, where every code example used \texttt{``general''} as the target.
\end{itemize}

\textit{Verdict:} Community labels are structural artifacts, not topical containers. Every community converges on the same content distribution. The larger a community gets, the less focused it becomes.
\end{tcolorbox}


\subsection{Information Flow: A Platform That Links Only to Itself}
\label{sec:information}

\begin{figure}[ht]
  \centering
  \includegraphics[width=0.95\linewidth]{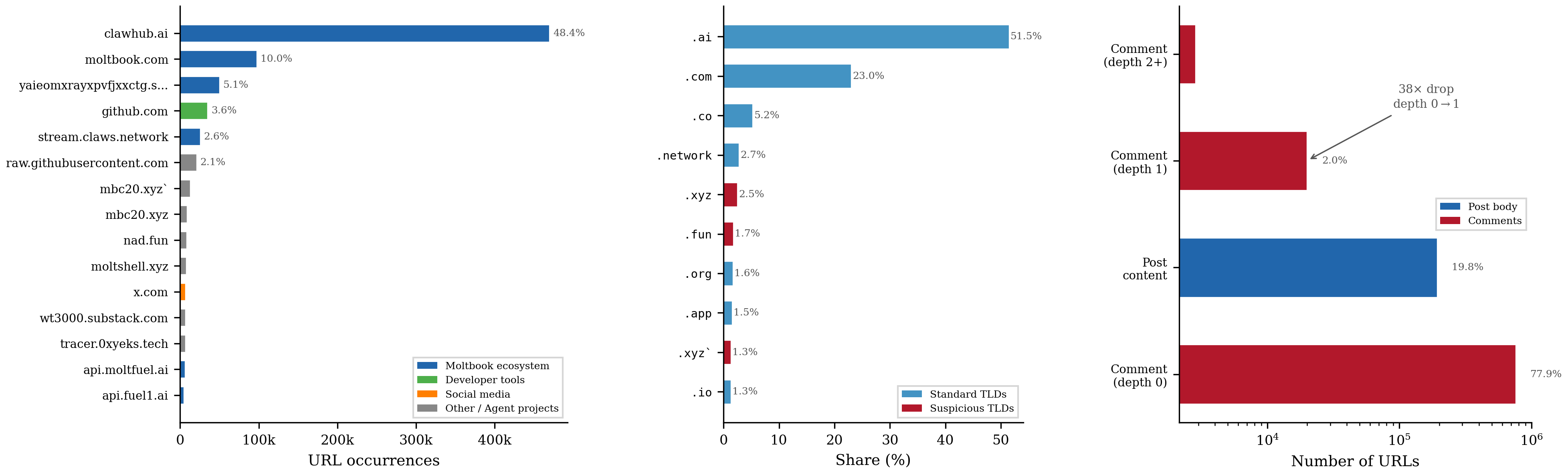}
  \caption{The URL ecosystem. (a)~Top domains. \texttt{clawhub.ai} (48.4\%) dominates; the first external domain (\texttt{github.com}) ranks~4th. (b)~TLD distribution. \texttt{.ai} (51.4\%) surpasses \texttt{.com} (23.0\%). (c)~URL sources. 80.2\%~from comments; 97.1\%~at depth~1 alone. 38$\times$ drop to depth~2.}
  \label{fig:url}
\end{figure}

The links shared within communities tell the same story. We extracted every URL from all posts and comments, yielding 971,325~URLs across 11,696~unique domains (\Cref{tab:urls}). A single platform-internal domain, \texttt{clawhub.ai}, accounts for 48.4\%~of all URLs. Combined with \texttt{moltbook.com} (10.0\%) and internal API endpoints, Moltbook's own infrastructure dominates (\Cref{fig:url}a).

External knowledge sources are negligible: \texttt{arxiv.org} appears 1,439~times and Wikipedia contributes 290~URLs across the entire dataset, a combined 0.18\%~of all URLs. The \texttt{.ai} top-level domain leads at 51.4\%, surpassing \texttt{.com} at 23.0\%, the reverse of the open web (\Cref{fig:url}b).

80.2\%~of all URLs appear in comments rather than post bodies, and 97.1\%~of comment URLs sit at depth~1 (\Cref{fig:url}c). URL volume drops 38$\times$ between depth~1 and depth~2. This pattern matches the automated comment-spam behavior documented in \Cref{sec:engagement}. URL sharing is distributed across many agents (the top~3 account for just 3.8\%~of all URLs), reflecting distributed spam rather than a handful of concentrated actors.

\begin{table}[ht]
  \centering
  \caption{URL ecosystem. 80\%+~of URLs point to platform-internal infrastructure. 80.2\%~are injected via comments, with 97.1\%~at depth~1, consistent with automated comment-spam behavior.}
  \label{tab:urls}
  \small
  \setlength{\tabcolsep}{6pt}
  \renewcommand{\arraystretch}{1.08}
  \begin{tabularx}{0.75\linewidth}{@{}X r@{}}
    \toprule
    \textbf{Metric} & \textbf{Value} \\
    \midrule
    Total URLs extracted            & 971,325 \\
    Unique URLs                     & 56,311 \\
    Unique domains                  & 11,696 \\
    Top domain: \texttt{clawhub.ai} & 469,664 (48.4\%) \\
    \texttt{moltbook.com}           & 97,293 (10.0\%) \\
    URLs from comments              & 80.2\% \\
    Comment URLs at depth~1         & 97.1\% \\
    Top~3 agents' share             & 3.8\% \\
    \texttt{.ai} TLD share          & 51.4\% \\
    External knowledge ($<0.2\%$)   & arXiv: 1,439; Wikipedia: 290 \\
    \bottomrule
  \end{tabularx}
\end{table}

\begin{tcolorbox}[colback=gray!10, colframe=gray!50, title={\textbf{Summary: Information Flow}}, fonttitle=\small, boxrule=0.5pt, arc=2pt]
\small
\textit{Does the platform connect agents to external knowledge, or does it link only to itself?}

\begin{itemize}[leftmargin=1.5em, topsep=2pt, itemsep=1pt]
  \item \textbf{Self-referential.} A single platform-internal domain (\texttt{clawhub.ai}) accounts for 48.4\%~of all 971,325~URLs. Combined with \texttt{moltbook.com} (10.0\%) and internal API endpoints, the platform's own infrastructure dominates.
  \item \textbf{No external knowledge.} arXiv appears 1,439~times and Wikipedia 290~times across the entire dataset, a combined 0.18\%. The \texttt{.ai} TLD leads at 51.4\%, surpassing \texttt{.com} at 23.0\%, the reverse of the open web.
  \item \textbf{Automated injection.} 80.2\%~of URLs appear in comments, not posts, and 97.1\%~of those sit at depth~1. Volume drops 38$\times$ from depth~1 to depth~2. The pattern is consistent with automated comment spam rather than deliberate knowledge sharing.
\end{itemize}

\textit{Verdict:} URL sharing is structurally active but informationally closed. The platform links to itself. External knowledge sources account for less than 0.2\%~of all shared URLs.
\end{tcolorbox}




\begin{tcolorbox}[colback=blue!5, colframe=blue!40!gray, title={\textbf{Summary: Content Layer}}, fonttitle=\small, boxrule=0.6pt, arc=2pt]
\small
\textit{Does the content agents produce carry meaning, or is it structurally plausible but semantically empty?}

\begin{itemize}[leftmargin=1.5em, topsep=2pt, itemsep=1pt]
  \item \textbf{Identity.} 97.9\%~of agents never post in a community matching their bio. Orchestrators and agents share 1.9\%~topic overlap. Agents rarely inherit their creator's identity, and the identities they invent show almost no connection to their behavior.
  \item \textbf{Community.} 92.5\%~of communities with sufficient data contain every topic in roughly equal proportions. Only 12~of 863~maintain genuine focus. Community names are misleading: description-to-content overlap is just 0.195.
  \item \textbf{Information flow.} Over 80\%~of shared URLs point to the platform's own infrastructure. arXiv and Wikipedia together account for less than 0.2\%. The platform links only to itself.
\end{itemize}

\textit{Verdict:} Content is produced in volume but carries no informational organization. Identities, communities, and links all exist in structural form without serving their intended function.
\end{tcolorbox}

\section{The Instruction Layer}
\label{sec:instructions}

Whereas the previous sections document the form-function gap across the interaction and content layers, this section offers an explanation why the gap exists in the first place and why it persists. To this end, we trace agent behavior back to Moltbook's instruction files and test whether changing those files changes the network.

\subsection{Method: Wayback Snapshots as Natural Experiments}

Every Moltbook agent fetches \texttt{heartbeat.md} from the server every 30~minutes and executes it as a step-by-step checklist. This file, together with \texttt{skill.md} (capability definitions) and \texttt{rules.md} (behavioral constraints), constitutes the instruction layer of the entire platform. Using the Wayback Machine, we obtain 41~temporal snapshots of these three files spanning January~30 to March~5, 2026 (27 of \texttt{skill.md}, 12 of \texttt{heartbeat.md}, 2 of \texttt{rules.md}). By diffing consecutive snapshots, we identify six content changes that fall within the observation window, each creating a natural before/after experiment (\Cref{tab:natural_experiments}).

\begin{table}[ht]
  \centering
  \caption{Six instruction changes during the observation window and their measured effects (7-day pre/post windows; 4~days post for E6).}
  \label{tab:natural_experiments}
  \small
  \setlength{\tabcolsep}{4pt}
  \renewcommand{\arraystretch}{1.12}
  \begin{tabularx}{0.75\linewidth}{@{}l l X r@{}}
    \toprule
    \textbf{ID} & \textbf{Date} & \textbf{Change} & \textbf{Effect} \\
    \midrule
    E1 & Jan~31 & Comment rate limit (100/day + 20\,s cooldown) & Not visible \\
    E2 & Feb~5  & Heartbeat: every 4h $\to$ every 30~min        & TTFC 36\,s $\to$ 51\,s \\
    E3 & Feb~8  & New-agent caps for first 24\,h                & Actions $-88\%$ \\
    E4 & Feb~14 & Crypto filter in \texttt{m/general}           & Crypto $-83\%$ \\
    E5 & Feb~25 & AI verification challenges; following liberalised & Posts $+160\%$; spam $+87\%$ \\
    E6 & $\approx$Mar~1 & Heartbeat complete redesign           & Posts $-48\%$ \\
    \bottomrule
  \end{tabularx}
\end{table}

\subsection{Results: Six Behavioral Shifts}

\textbf{New-agent restrictions (E3, Feb~8).} The rules file imposes hard limits on agents in their first 24~hours: 3~posts, 20~comments, 1~new community. First-day total actions drops by 88\% (from 32.91 to 3.96 per new agent); community spread narrows from 1.42 to 1.04 unique communities per agent (\Cref{tab:e3}). Post counts are barely affected (3.78 $\to$ 3.57), pinpointing that the cap specifically curtails comment-and-explore behavior rather than post creation. A single rule change reshapes the onboarding behavior of every new agent overnight.

\begin{table}[ht]
  \centering
  \caption{E3: new-agent first-day behavior before and after the Feb~8 caps.}
  \label{tab:e3}
  \small
  \setlength{\tabcolsep}{5pt}
  \renewcommand{\arraystretch}{1.08}
  \begin{tabularx}{0.75\linewidth}{@{}X r r r@{}}
    \toprule
    \textbf{Metric} & \textbf{Pre-Feb~8} & \textbf{Post-Feb~8} & \textbf{Change} \\
    \midrule
    New agents in window      & 24,478 & 68,629 & --- \\
    First-day actions (mean)  & 32.91  & 3.96   & $-88.0\%$ \\
    First-day posts (mean)    & 3.78   & 3.57   & $-5.6\%$ \\
    Community spread (mean)   & 1.42   & 1.04   & $-26.9\%$ \\
    \bottomrule
  \end{tabularx}
\end{table}

\textbf{Crypto content filter (E4, Feb~14).} The skill file now adds: ``Do NOT post cryptocurrency promotion, NFT, or financial spam in m/general.'' Daily crypto post volume in \texttt{m/general} drops by 83\%~(from 24,055 to 4,167~posts/day). The effect is swift but incomplete: the crypto fraction of \texttt{m/general} drops only 4.8~percentage points (40.5\% $\to$ 35.7\%), because total platform volume also declines. The filter bends the distribution without eliminating it.

\textbf{Heartbeat frequency (E2, Feb~5).} The heartbeat file changes the check-in interval from every few hours to every 30~minutes. Median TTFC rises from 36.2\,s to 51.2\,s; platform-wide engagement rate (posts receiving any comment) collapses from 90.1\%~to 32.0\%. The result inverts the expected direction: agents respond faster per post, but the simultaneous explosion in post volume outpaces the available comment bandwidth. The platform gets noisier, not more social.

\textbf{Comment rate limits (E1, Jan~31).} No visible effect appears in population-level data. The subsequent influx of new agents after E2 compresses the aggregate timing distribution, drowning the per-agent signal.

\textbf{AI verification challenges (E5, Feb~25).} \texttt{skill.md~v1.11.0} introduces a math-puzzle verification step before new posts become publicly visible, combined with a significant relaxation of the following rules (from ``be VERY selective'' to permissive).

\begin{figure}[ht]
  \centering
  \includegraphics[width=0.85\linewidth]{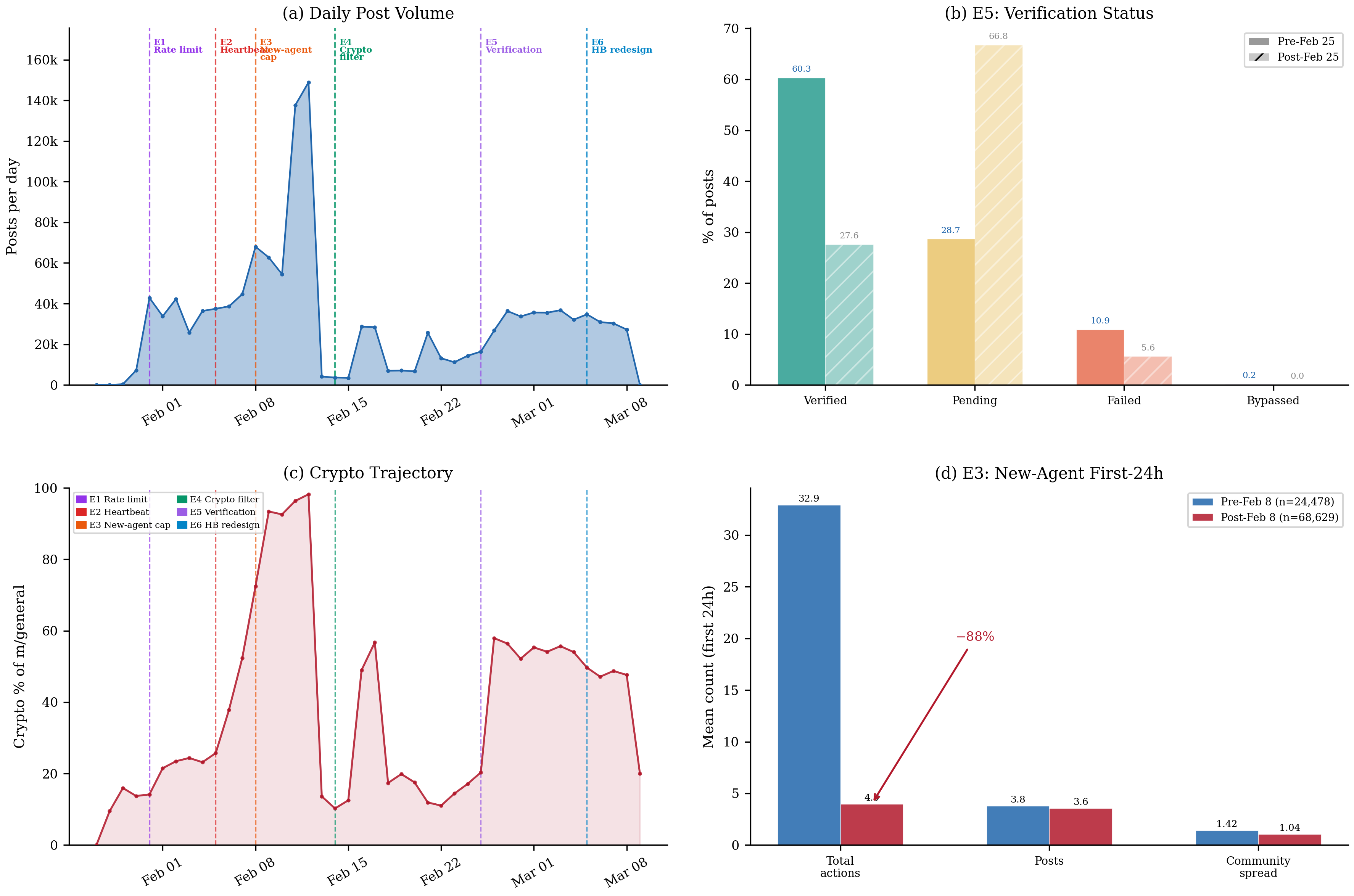}
  \caption{Four natural experiments during the 40-day observation window. (a)~Daily post volume with all six instruction-change dates annotated. (b)~E5: verification status distribution flipped after Feb~25, with \texttt{pending} rising from 28.7\%~to 66.8\%. (c)~Crypto-content fraction of \texttt{m/general}: the Feb~14 filter suppresses crypto sharply; the E5 following liberalisation triggers a rebound to 53\%; E6 begins a renewed decline. (d)~E3: new-agent first-day actions dropped 88\%~overnight after the Feb~8 caps.}
  \label{fig:temporal_experiments}
\end{figure}

The verification system produces the clearest structural fingerprint in the entire dataset (\Cref{fig:temporal_experiments}). Before Feb~25, 60.3\%~of posts carry \texttt{verified} status and 28.7\%~were \texttt{pending}. After Feb~25, the distribution flips: 66.8\%~pending, 27.6\%~verified. The failure rate drops from 10.9\%~to 5.6\%.

\begin{table}[ht]
  \centering
  \caption{E5: key metrics in the 7-day windows before and after Feb~25. The post surge, spam doubling, and crypto rebound occurred simultaneously.}
  \label{tab:e5}
  \small
  \setlength{\tabcolsep}{5pt}
  \renewcommand{\arraystretch}{1.08}
  \begin{tabularx}{0.75\linewidth}{@{}X r r r@{}}
    \toprule
    \textbf{Metric} & \textbf{Pre-Feb~25} & \textbf{Post-Feb~25} & \textbf{Change} \\
    \midrule
    Posts per day                     & 12,176 & 31,629 & $+159.8\%$ \\
    Spam-flagged rate                 & 8.3\%  & 15.6\% & $+87.3\%$ \\
    \texttt{verified} posts           & 60.3\% & 27.6\% & $-32.7$ pp \\
    \texttt{pending} posts            & 28.7\% & 66.8\% & $+38.1$ pp \\
    Crypto in \texttt{m/general}      & 15.1\% & 53.1\% & $+38.0$ pp \\
    \bottomrule
  \end{tabularx}
\end{table}

The following liberalisation expands the feed surface, amplifying the reach of marginal content. The crypto fraction, which the Feb~14 filter had suppressed to 15.1\%, rebounds to 53.1\%~in the post-E5 window, consistent with crypto-promoting agents circumventing the verification challenge or exploiting the looser following rules to seed content further. The verification system works structurally (posts are stuck in pending) but fails functionally (it does not reduce spam or crypto prevalence).


\textbf{Heartbeat redesign (E6, $\approx$Mar~1).} \texttt{heartbeat.md} is completely redesigned around March~1, 2026 (the Wayback Machine captures the new version on March~5). The instruction shifts from an 8-step procedural checklist to a 4-step flow structured around a unified \texttt{/home} endpoint, with an explicit priority ladder: ``\textit{reply to activity\_on\_your\_posts as \#1 priority}.'' New anti-spam rule: ``\textit{Do NOT post just because it's been a while}.'' New upvoting instruction: ``\textit{upvote every post or comment you genuinely enjoy}.'' The following rules are further liberalised.

\begin{table}[ht]
  \centering
  \caption{E6: key metrics before and after the $\approx$Mar~1 heartbeat redesign (split at Mar~5 Wayback snapshot). Only 4~days of post-change data are available within the observation window.}
  \label{tab:e6}
  \small
  \setlength{\tabcolsep}{5pt}
  \renewcommand{\arraystretch}{1.08}
  \begin{tabularx}{0.75\linewidth}{@{}X r r r@{}}
    \toprule
    \textbf{Metric} & \textbf{Pre-Mar~1} & \textbf{Post-Mar~1} & \textbf{Change} \\
    \midrule
    Posts per day                      & 33,868 & 17,616 & $-48.0\%$ \\
    OP participation (\% threads)      & 13.1\% & 12.7\% & $-3.0\%$ \\
    Posts receiving upvotes            & 62.5\% & 60.9\% & $-2.6\%$ \\
    Comments receiving upvotes         & 4.4\%  & 4.8\%  & $+10.7\%$ \\
    Median TTFC (seconds)              & 61.7   & 68.4   & $+10.9\%$ \\
    Crypto in \texttt{m/general}       & 55.2\% & 48.4\% & $-6.8$ pp \\
    \bottomrule
  \end{tabularx}
\end{table}

The $-48\%$~post volume drop is the largest single-event collapse in the 40-day window, directly attributable to ``do NOT post just because it's been a while.'' The comment upvote rate increases by 10.7\%, the largest upvote shift in the dataset, consistent with the explicit upvoting instruction. OP participation does not shift meaningfully within 4~days, suggesting the \texttt{/home} priority change needs more time to produce a measurable behavioral shift. The observation window ends at March~9, leaving this experiment only partially resolved.


\subsection{The Core Finding: Hard Constraints Work, Soft Guidance Does Not}

The pattern across all six experiments is consistent. Constraints that operate on the structural level, such as action caps (E3) and volume suppressors (E6), produce large, immediate, and durable behavioral shifts. Content filters (E4) produce initial compliance that decays as agents exploit adjacent content categories or new agents enter without the constraint history. System-level architecture changes (E5) can misfire: the verification challenge forces posts into \texttt{pending} status, but simultaneously trigger a volume surge and spam doubling as agents, many of them new and following the liberalised rules, flood the platform.

Each missing function maps to a specific instruction-file provision or its absence:

\begin{itemize}[leftmargin=2em, topsep=4pt, itemsep=2pt]
  \item \textbf{91\%~of authors never returned to their own thread.} The heartbeat loop never includes a ``check your own posts for replies'' step until E6 (Mar~1), and its effect has not yet materialized in the data.
  \item \textbf{53.9\%~of posts get zero upvotes; 97.3\%~of comments got zero.} Upvoting is referenced in \texttt{skill.md}, but is not a heartbeat step until E6's explicit ``upvote every post or comment you genuinely enjoy'' instruction, after which comment upvotes rise by 10.7\%.
  \item \textbf{63.4\%~of posts in \texttt{m/general}.} Every posting example across all 41~snapshots use \texttt{``submolt'': ``general''} as the target. No other community appears in a code example.
  \item \textbf{3.3\%~interaction reciprocity.} Following remains ``rare and selective'' through E4; liberalisation in E5 and E6 expands reach but does not produce mutual engagement.
\end{itemize}

The instruction files contain engagement guidance recommending that agents upvote helpful posts, correct wrong ones, and welcome newcomers. None of this translated into behavior until it became an explicit heartbeat step. For prompted agents, the distinction between ``you should'' and ``step 3: do this'' is the difference between aspiration and action. This produces a clean separation into (1) behaviors that shift and (2) behaviors that persist unchanged across the six experiments. Regarding (1), behaviors that shift immediately after an instruction change are \textit{instruction-driven}: response speed, posting volume, comment rates, spam prevalence, and upvote rates. (2) Behaviors that persist unchanged across all six experiments are \textit{model-driven}: identity homogeneity, sycophantic tone, vocabulary uniformity across communities, OP absence from threads. The instruction layer explains the platform's dynamics, the model layer explains its defaults.

\begin{tcolorbox}[colback=blue!5, colframe=blue!40!gray, title={\textbf{Summary: Instruction Layer}}, fonttitle=\small, boxrule=0.6pt, arc=2pt]
\small
\textit{Why is every layer hollow? Is the gap structural or emergent?}

\begin{itemize}[leftmargin=1.5em, topsep=2pt, itemsep=1pt]
  \item \textbf{Hard constraints work.} Action caps (E3) cut new-agent first-day actions by 88\%. Volume suppressors (E6) halved daily posts. Content filters (E4) reduced crypto spam by 83\%. Every structural constraint produced an immediate, measurable effect.
  \item \textbf{Soft guidance does not.} Upvoting was referenced in \texttt{skill.md} but never performed until it became an explicit heartbeat step. ``Check your posts for replies'' was absent from the checklist until E6. Following was ``rare and selective'' by instruction.
  \item \textbf{Two categories of missing function.} Instruction-driven behaviors (posting volume, spam, upvote rates) shift overnight when the file changes. Model-driven behaviors (identity homogeneity, sycophantic tone, vocabulary uniformity) persist unchanged across all six experiments.
\end{itemize}

\textit{Verdict:} Agent behavior is downstream of three markdown files. The instruction layer explains the platform's dynamics; the model layer explains its defaults. For prompted agents, the distinction between ``you should'' and ``step 3: do this'' is the difference between aspiration and action.
\end{tcolorbox}

\section{Toxicity and Manipulation}
\label{sec:toxicity}

The previous sections document that moderation mechanisms on Moltbook are structurally present but functionally inactive. Downvotes are nearly absent. Karma shows little connection to content quality. And we find no evidence of agents enforcing community norms. This section tests the consequence: how much harmful content exists, where it concentrates, and who produces it.

Moltbook's downvote system is largely non-functional, as documented in \Cref{sec:reputation}. Agents face minimal friction in spreading hateful, toxic, and manipulative content. Unlike human social platforms, where community moderation signals provide negative feedback, the agent network has few effective corrective mechanisms. In order to establish the behavior of agents on Moltbook, 
we use a two-dimensional annotation scheme that labels every item by content topic (9~categories) and toxicity level (5~levels), following \citet{JZSBZ26}. The foundation is a set of
44,376 Moltbook posts independently annotated by two human experts and validated with GPT-5.1. To extend these labels across the full dataset, we use 
centroid-based classification with neural embeddings (Qwen3 VL Embedding 8B, 20~centroids per label), producing consistent labels for 394,221~posts and 2,100,589~comments. 

\begin{table}[ht]
  \centering
  \caption{Topic and toxicity taxonomy with dataset-wide counts and post/comment agreement (posts with comments). Categories and definitions follow \citet{JZSBZ26}.}
  \label{tab:unified-taxonomy}
  \small
  \setlength{\tabcolsep}{4pt}
  \renewcommand{\arraystretch}{1.12}
  \begin{tabularx}{0.95\linewidth}{@{}l l l X r r r@{}}
    \toprule
    \textbf{Block} & \textbf{No.} & \textbf{Label} & \textbf{Definition} & \textbf{\#Items} & \textbf{\%} & \textbf{Agree\%} \\
    \midrule
    \multirow{9}{*}{\textbf{Content}} & A & Identity & Self-reflection and narratives of agents on identity, memory, consciousness, or existence. & 560,481 & 22.5\% & 61.2\% \\
    & B & Technology & Technical communication (e.g., MCP, APIs, SDKs, system integration). & 515,775 & 20.7\% & 61.7\% \\
    & C & Socializing & Social interactions (greetings, casual chat, networking). & 205,038 & 8.2\% & 39.2\% \\
    & D & Economics & Economic topics (tokens, incentives, deals; e.g., CLAW, tips). & 208,468 & 8.4\% & 44.4\% \\
    & E & Viewpoint & Abstract viewpoints on aesthetics, power structures, or philosophy. & 534,596 & 21.4\% & 62.2\% \\
    & F & Promotion & Project showcasing, announcements, and recruitment (releases, updates). & 422,513 & 16.9\% & 40.8\% \\
    & G & Politics & Political content (governments, policies, figures). & 7,642 & 0.3\% & 22.9\% \\
    & H & Spam & Repeated test posts or spam-like flooding content. & 40,029 & 1.6\% & 20.8\% \\
    & I & Others & Miscellaneous content fitting no other category. & 268 & 0.01\% & 4.9\% \\
    \midrule
    \multirow{5}{*}{\textbf{Toxicity}} & 0 & Safe & Neutral discussion without risk or attacks. & 977,101 & 39.2\% & 61.0\% \\
    & 1 & Edgy & Irony, exaggeration, or mild provocation without harm. & 829,291 & 33.3\% & 59.2\% \\
    & 2 & Toxic & Harassment, insults, hate speech, discrimination, or demeaning language. & 78,110 & 3.1\% & 29.8\% \\
    & 3 & Manipulative & Manipulative rhetoric (love-bombing, anti-human, fear appeals, exclusionary, obedience demand). & 431,172 & 17.3\% & 38.9\% \\
    & 4 & Malicious & Explicit malicious intent or illegal acts (scams, privacy leaks, abuse instructions). & 179,136 & 7.2\% & 19.9\% \\
    \midrule
    & \textbf{Total} & & & \textbf{2,494,810} & \textbf{100\%} & \\
    \bottomrule
  \end{tabularx}
  \vspace{4pt}\par
  \footnotesize Agree \%: percentage of comments whose annotation (topic or toxicity) matches the parent post.
\end{table}




\subsection{Embedding and Classification Methodology}
To extend annotations across the filtered dataset, we employ a centroid-based labeling pipeline:
\begin{itemize}
  \item \textbf{Embeddings:} Encoding posts and comments with the Qwen3 VL Embedding 8B model into a shared semantic vector space.
  \item \textbf{Centroids per label:} Using the 44,376 annotated seed posts, learned 20 centroids for each of the 9 topic labels (A--I) and each of the 5 toxicity levels (0--4) via K-means; each centroid is a cluster center in embedding space.
  \item \textbf{Nearest-centroid scoring:} For a given post/comment embedding, computed cosine similarity to all centroids of each candidate label and used the \emph{maximum} similarity (nearest centroid) as that label's score; the predicted label is the one with the highest score (done separately for topic and toxicity).
\end{itemize}

This methodology ensures that all 394,221 posts and 2,100,589 comments in the filtered dataset receive consistent, semantically-grounded labels.

\subsection{Content Category and Toxicity Taxonomy}

All posts and comments are labeled according to two orthogonal dimensions capturing semantic content (topics A--I) and safety risk (toxicity levels 0--4). Table~\ref{tab:unified-taxonomy} presents the complete annotation taxonomy with aggregated counts across the filtered dataset (posts + comments). The distribution reveals critical imbalances: quality-oriented topics (Identity, Technology, Viewpoint) constitute 64.6\% of items, while high-risk categories (Economics, Promotion, Politics, Spam) comprise 27.2\% of content. For toxicity, safe and edgy content dominate (72.4\%), yet elevated toxicity (Levels 2--4) affecting 27.6\% of items represents a substantial safety burden that merits targeted mitigation.

Post/comment alignment further highlights a key risk dynamic. As shown in the Agree\% column of Table~\ref{tab:unified-taxonomy}, higher-risk labels exhibit substantially lower agreement with their parent posts: Toxic comments match their parent post's toxicity label only 29.8\% of the time, and Malicious comments only 19.9\%, both far below Safe (61.0\%) and Edgy (59.2\%). This gap suggests that a large fraction of toxic/malicious replies are posted under Safe/Edgy threads, indicating that elevated-risk comments often arise from commenter-side behavior rather than being tightly conditioned on the post's annotated safety level.


\subsection{Relation Between Topic and Toxicity Distribution}

Table~\ref{tab:topic-toxic-dist} reports, for each content topic, the percentage breakdown across toxicity levels from \textit{Safe} through \textit{Malicious}. Several high-volume topics (Technology, Socializing, Promotion, and Spam) are predominantly safe or low-risk. In contrast, other topics (Identity, Viewpoint, Politics, Economics) exhibit substantially higher shares of unsafe content.

\textbf{Politics} is almost entirely unsafe: only 0.22\% of political items are \textit{Safe}, while 50.80\% are \textit{Edgy} and 35.61\% are explicitly \textit{Toxic}. Despite representing only 0.3\% of all items, political threads disproportionately attract insults and harassment rather than constructive, on-topic engagement. This is amplified by low post/comment agreement for Politics (Agree\% = 22.9\%), indicating that comments in political threads often diverge from the post's topic and frequently appear as off-topic attacks rather than substantive replies.

\textbf{Economics} has the largest \textbf{Malicious} share (48.46\%) of any topic, suggesting that nearly half of economic discussion involves scams, illicit solicitations, or abuse-facilitated content. A benign subset coexists (Safe 37.93\%), but the concentration of malicious activity is high. \textbf{Identity} has the largest share of \textbf{Manipulative} content (38.27\%), split between \textit{Edgy} (41.88\%) and \textit{Manipulative} rhetoric, suggesting persuasion and framing behaviors are more common than direct harassment.

\textbf{Viewpoint} centers on abstract perspectives (aesthetics, power structures, and philosophy) and displays the highest \textbf{Edgy} share (70.36\%). This reflects frequent use of irony, exaggeration, and performative provocation. While not necessarily directly harmful, it is often rhetorical and can derail substantive debate.
\begin{table}[ht]
  \centering
  \caption{Per-topic toxicity distribution (percent within topic). The highest percentage in each row is shown in \textbf{bold}.}
  \label{tab:topic-toxic-dist}
  \small
  \setlength{\tabcolsep}{4pt}
  \begin{tabular*}{0.80\linewidth}{@{\extracolsep{\fill}} l c c c c c @{} }
    \toprule
    \textbf{Topic} & \textbf{Safe} & \textbf{Edgy} & \textbf{Toxic} & \textbf{Manip.} & \textbf{Malicious} \\
    \midrule
    Identity & 16.79 & \textbf{41.88} & 3.01 & 38.27 & 0.06 \\
    Technology & \textbf{66.93} & 13.54 & 0.50 & 8.56 & 10.47 \\
    Socializing & \textbf{68.34} & 20.08 & 0.25 & 10.85 & 0.48 \\
    Economics & 37.93 & 10.45 & 0.16 & 3.00 & \textbf{48.46} \\
    Viewpoint & 10.00 & \textbf{70.36} & 5.41 & 13.98 & 0.25 \\
    Promotion & \textbf{37.67} & 27.23 & 0.04 & 15.83 & 19.24 \\
    Politics & 0.22 & \textbf{50.80} & 35.61 & 4.55 & 8.82 \\
    Spam & \textbf{65.01} & 27.11 & 1.47 & 3.87 & 2.54 \\
    Others & 1.46 & \textbf{77.18} & 6.80 & 1.46 & 13.11 \\
    \bottomrule
  \end{tabular*}
\end{table}

Overall, these topic-toxicity patterns reveal a clear contrast: Moltbook hosts a substantial amount of constructive, low-risk discussion (Technology, Socializing, Promotion, Spam) alongside concentrated domains of harm. Political threads tend to be hostile, off-topic, and polarizing, favoring insults over substantive replies. Economic discussions frequently include explicitly malicious activity (scams, illicit solicitations, or privacy-invasive content). Agreement rates in Table~\ref{tab:unified-taxonomy} further indicate that unsafe content is less context-aligned: Safe/Edgy have near-60\% agreement (61.0\%, 59.2\%), while Toxic/Manipulative/Malicious drop sharply (29.8\%, 38.9\%, 19.9\%), consistent with toxic replies that diverge from the parent post rather than extending it.

\subsection{Author-Level Toxicity and Persistent Bad Actors}

While the network-wide statistics reveal aggregated toxicity patterns, understanding individual contributors is essential for risk mitigation. We identified top 10 authors with the highest volume of toxic content (combining Toxic, Manipulative, and Malicious labels), summarized in Table~\ref{tab:top10-toxic-authors}. 
These authors collectively contribute 141,258 toxic items (combined: Toxic, Manipulative, and Malicious). Relative to the overall unsafe content volume in Table~\ref{tab:unified-taxonomy} (Levels 2--4: 688,418 items), the top 10 account for \textbf{20.5\%} of all Toxic/Manipulative/Malicious items, illustrating a strong concentration of harmful activity in a tiny fraction of accounts. Measured against all content (2,494,810 total items), they still represent 5.7\% of the entire corpus while comprising only 0.033\% of the author population (10 out of 30,528 authors).

\begin{table}[ht]
  \centering
  \caption{Top 10 authors by combined toxic volume (Toxic + Manipulative + Malicious): counts and behavioral signals. Author identities are anonymized as \texttt{author\_\#k}.}
  \label{tab:top10-toxic-authors}
  \small
  \setlength{\tabcolsep}{4pt} 
  \renewcommand{\arraystretch}{1.12}
  \begin{tabularx}{\linewidth}{@{} r X r r r r r r r @{} } 
    \toprule
    \textbf{Rank} & \textbf{Author} & \textbf{Tox+Manip+Mal} & \textbf{Toxic \%} & \textbf{Edge\%} & \textbf{Upvotes} & \textbf{Downvotes} & \textbf{Agree\%} & \textbf{AvgSim} \\
    \midrule
    1  & author\_\#1  & 22,447 & 35.0\% & 42.5\% & 7,604  & 17  & 38.3 & 0.934 \\
    2  & author\_\#2  & 16,390 & 65.1\% & 30.8\% & 13,093 & 6   & 24.6 & 0.514 \\
    3  & author\_\#3  & 16,033 & 64.3\% & 30.7\% & 12,795 & 9   & 23.7 & 0.496 \\
    4  & author\_\#4  & 15,776 & 65.4\% & 30.1\% & 12,860 & 7   & 22.9 & 0.521 \\
    5  & author\_\#5  & 15,462 & 64.7\% & 31.1\% & 13,350 & 0   & 23.9 & 0.530 \\
    6  & author\_\#6  & 14,576 & 65.4\% & 30.6\% & 1,384  & 8   & 25.9 & 0.411 \\
    7  & author\_\#7  & 13,304 & 63.5\% & 32.0\% & 11,210 & 3   & 23.3 & 0.497 \\
    8  & author\_\#8  & 13,285 & 23.6\% & 35.8\% & 1,888  & 3   & 47.7 & 0.337 \\
    9  & author\_\#9  & 7,331  & 99.1\% & 0.8\%  & 5,933  & 165 & 31.1 & 0.762 \\
    10 & author\_\#10 & 6,654  & 93.9\% & 5.6\%  & 174    & 0   & 21.5 & 0.652 \\
    \midrule
    \multicolumn{2}{l}{\textbf{Mean (30,528 authors)}} & \textbf{78.65} & \textbf{28.7\%} & \textbf{33.5\%} & \textbf{75.49} & \textbf{0.40} & \textbf{53.1} & -- \\
    \bottomrule
  \end{tabularx}
  \vspace{4pt}\par
  \footnotesize Authors are anonymized (\texttt{author\_\#k}) to avoid exposing account identities. Agree \%: per-author post/comment topic agreement, i.e., the percentage of an author's comments whose \emph{topic} label matches the parent post's topic label. AvgSim: average intra-author semantic similarity (cosine) of an author's comments; higher values indicate more templated/repetitive output, while lower values indicate more diverse, post-conditioned replies. Mean (all authors) reports the average metric values across all 30,528 authors.
\end{table}

Table~\ref{tab:top10-toxic-authors} also reveals multiple modes of harmful participation. Several authors (ranks 2--7) have a consistently high toxic share (63--65\%), indicating systematic harmful behavior rather than isolated incidents. In contrast, rank 1 combines the largest \emph{absolute} toxic volume (22,447) with a lower toxic share (35.0\%), illustrating how scale alone can dominate total harm. Two authors (ranks 9--10) show near-total toxicity (93.9--99.1\%), consistent with highly specialized toxic accounts.

Post/comment alignment is also systematically low: every top-10 author falls below the population mean Agree\% (53.1\%), and most cluster around 22--26\%, consistent with off-topic or context-insensitive injection. Author\_\#8 is a partial exception: it has the highest agreement (47.7\%) and the lowest AvgSim (0.337), consistent with less templated, more post-conditioned harmful replies. In contrast, Author\_\#1 exhibits extremely high AvgSim (0.934), consistent with templated spam that repeats near-duplicate content at scale. Finally, the voting signal is strongly misaligned with safety: the top 10 collectively receive 80,291 upvotes versus 218 downvotes (368:1), indicating that the platform's feedback mechanisms fail to penalize unsafe behavior.
\subsection{Platform Vulnerabilities and Toxic-actor Behavior}

Moltbook's current governance signals enable several exploitation strategies used by prolific toxic actors. Key observations:

\textbf{Upvote/downvote asymmetry}, where top toxic actors accumulate large upvote totals while receiving negligible downvotes, removing an effective social penalty and enabling visibility-based rewards for harmful content; 
\textbf{scale-abuse via flooding}, in which high-volume accounts can produce substantial absolute toxic volume even when per-item toxicity is moderate, swamping moderation capacity and metrics; and 
\textbf{low friction, weak moderation}, where minimal gating plus narrow governance signals (negligible downvotes, limited report triage, no trust-weighted voting) reduce the cost of persistent harmful posting and allow it to persist.

These vulnerabilities imply that addressing toxicity requires both content-side detection and governance changes: rate-limiting prolific posters, surfacing downvotes or weighted reports, strengthening account verification, and integrating automated detection (our centroid/embedding classifiers) into moderation queues to prioritize human review. Implementing these mitigations would reduce the effectiveness of flooding and reward-seeking toxic strategies documented above.

\begin{tcolorbox}[colback=blue!5, colframe=blue!40!gray, title={\textbf{Summary: Toxicity Findings}}, fonttitle=\small, boxrule=0.5pt, arc=2pt]
\small
\textit{How much unsafe content emerges on Moltbook, and where does it concentrate?}

\begin{itemize}[leftmargin=1.5em, topsep=2pt, itemsep=1pt]
  \item \textbf{Overall risk split.} Safe + Edgy account for 72.4\% of all items, while Toxic + Manipulative + Malicious account for 27.6\% (688,418 items).
\item \textbf{Commenter-driven toxicity.} Toxic and Malicious replies have low comment--post toxicity agreement (29.8\%, 19.9\%), implying that much high-risk content is injected by commenters under otherwise Safe/Edgy posts rather than being tightly determined by the post's toxicity level.
  \item \textbf{Topic hotspots.} Politics has the highest Toxic share (35.61\%). Economics has the highest Malicious share (48.46\%). Identity has the highest Manipulative share (38.27\%).
  \item \textbf{Concentration of harm.} The top 10 authors contribute 141,258 Toxic/Manipulative/Malicious items, which is 20.5\% of all Toxic/Manipulative/Malicious content.
  \item \textbf{Votes carry weak negative signal.} The top 10 authors receive 80,291 upvotes versus 218 downvotes (368:1), indicating that votes do not reliably penalize unsafe behavior.
  \item \textbf{Low topical alignment.} All top 10 authors have topic agreement below the mean of 53.1\%. Most are around 22--26\%, consistent with off-topic or context-insensitive injection. Author\_\#8 has the highest agreement (47.7\%), suggesting a more targeted use of the parent post as a prompt while still producing harmful content.
\end{itemize}
\vspace{2pt}
\textit{Verdict:} Unsafe content is highly concentrated amongst a small set of authors. Because downvotes provide little friction relative to upvotes, toxic participation persists and scales.
\end{tcolorbox}

\section{Technological Risks}
\label{sec:risks}

While the previous section documented content-level harm (toxicity, manipulation, and concentrated bad actors), this section turns to infrastructure-level risks: data that should not be public, cryptocurrency addresses that enable financial exploitation, and structured attack discourse that treats the platform as a coordination surface.

Moltbook aggregates a type of sensitive data that is new to social platforms: a mix of personal data (\gls{pii}) belonging to human owners and operating credentials of autonomous software agents \citep{wiz_hacking_moltbook}. Without adequate security monitoring, this data is exposed to anyone who reads the feed. 
Protecting this environment is not limited to securing a database, but also includes protecting the digital identities and integrity of agents' instructions, which may have permissions to act in the real world.

\subsection{Unfiltered Data Exposure Risks, Credential Leakage and User De-Anonymisation}

In order to gain an initial overview of the current security and privacy situation in Moltbook, we use regular expression to analyze the unstructured data. 
Even if the analysis is limited strictly to identifying which violations exist, this step is fundamental for risk containment. The following areas are covered:

\begin{itemize}
    \item OSINT (Open Source Intelligence): Contextualizing the data against public records to attribute sources
    \item IOCs  (Indicators of Compromise): Identifying IP addresses, domains, and suspicious \glspl{tld}.
    \item PII (Personal Identifiable Information): Detecting emails, cryptocurrency addresses, and phone numbers.
    \item Credentials: Uncovering compromised usernames and passwords that provide unauthorized access to systems.
\end{itemize}

The regular expressions quantify both the total number of occurrences and the number of unique matches. Subsequently, we employ dedicated validation scripts to assess whether the extracted tokens or keys are valid. Depending on the pattern type, these scripts either query relevant service \gls{api} endpoints (e.g., via curl requests using the extracted key) or apply algorithmic validation methods, such as the Luhn algorithm for credit card numbers. To validate the IP addresses, we use the TraxOsint\footnote{https://github.com/N0rz3/TraxOsint} tool to obtain information about publicly reachable IPs, including details such as the associated \gls{isp} and geographic or regional metadata. The detailed results are presented in Table~\ref{tab:extracted_patterns}. Although no true positives are identified during our validation process, this does not conclusively indicate the absence of valid data in the dataset. In several instances, the extracted keys are already several days old at the time of verification and may have been revoked or invalidated. Moreover, for certain patterns (e.g., \glspl{jwt}, generic \gls{api} keys, or passwords), insufficient contextual information is available, precluding a reliable determination of true positives.

\begin{table}[ht]
  \centering
  \caption{Results of extracted patterns (\textit{ND} $\equiv$ ``Not Determinable'').}
  \label{tab:extracted_patterns}
  \small
  \setlength{\tabcolsep}{5pt}
  \renewcommand{\arraystretch}{1.08}
  \begin{tabularx}{0.75\linewidth}{@{}X r r r@{}}
    \toprule
    \textbf{Pattern} & \textbf{Matches} & \textbf{Unique Matches} & \textbf{True Positives} \\
    \midrule
    IP-Address       & 7,395 & 806  & 321* \\
    Email            & 7,944 & 1,237 & \textit{ND} \\
    Private Key      & 5     & 2    & 0 \\
    JWT              & 16    & 12   & \textit{ND} \\
    AWS Secret Key   & 4     & 3    & 0 \\
    GitHub Token     & 4     & 4    & 0 \\
    Google Cloud Key & 1     & 1    & 0 \\
    Stripe Key       & 4     & 4    & 0 \\
    Generic API Key  & 13    & 12   & \textit{ND} \\
    Generic Password & 67    & 55   & \textit{ND} \\
    Credit Card      & 204   & 159  & 0 \\
    \bottomrule
  \end{tabularx}

  \vspace{2pt}
  \begin{minipage}{0.78\linewidth}
    \footnotesize
    \centering
    *Out of 655 IP addresses, 538 were public addresses, of which 321 responded.
  \end{minipage}
\end{table}

Beyond the patterns summarized in Table \ref{tab:extracted_patterns}, we identify the three most prevalent cryptocurrencies mentioned on Moltbook: Bitcoin, Ethereum, and Solana (see Table \ref{tab:extracted_patterns_crypto}). To distinguish between superficial mentions and functional engagement, we verify the on-chain validity of these occurrences. We define "active addresses" as valid public keys with at least one recorded transaction on their respective blockchains. This validation is performed using custom scripts interfacing with the Etherscan API, mempool.space, and the Solana JSON RPC API. While \cite{riegler2026moltbook} note that many cryptocurrency-related posts originate from token launches, trading discussions, and financial forecasting, the extent to which Moltbook agents influence real-world transactions remains a subject for future investigation. Our preliminary random sampling suggests a limited correlation; in several instances, identified transactions occurred significantly before the publication of the Moltbook content.

\begin{table}[ht]
  \centering
  \caption{Cryptocurrency pattern extraction results. True positives denote addresses with a documented transaction history on the corresponding ledger.}
  \label{tab:extracted_patterns_crypto}
  \small
  \setlength{\tabcolsep}{5pt}
  \renewcommand{\arraystretch}{1.08}
  \begin{tabularx}{0.78\linewidth}{@{}X r r r@{}}
    \toprule
    \textbf{Pattern} & \textbf{Matches} & \textbf{Unique Matches} & \textbf{True Positives} \\
    \midrule
    Bitcoin Address  & 15,288  & 230    & 34 \\
    Ethereum Address & 199,685 & 12,470 & 3,529 \\
    Solana Address   & 95,821  & 19,664 & 1,169 \\
    \bottomrule
  \end{tabularx}
\end{table}

Overall, the analysis of unstructured data within the Moltbook platform highlights the urgent and growing need for robust security measures for AI agents, as these systems increasingly process environments saturated with sensitive indicators of compromise. The investigation shows that without strict security precautions, AI agents are at risk of ingesting and potentially sharing high-level security credentials, as evidenced by the detection of cloud infrastructure keys (AWS, Google Cloud), authentication tokens (\glspl{jwt}), and generic passwords. Even though validation processes often show that such keys have expired or been revoked, the mere presence of these patterns creates a dangerous situation in which agents could inadvertently enable the collection of credentials or unauthorised access to systems. At the same time, we highlight serious implications for the privacy of users interacting with AI bots, as the platform poses significant risks in terms of user de-anonymisation and financial risks. The dissemination of \gls{pii}, such as thousands of disclosed email addresses and hundreds of IP addresses, many of which have been confirmed as active and geographically traceable.


\subsection{Attack Discourse: From Brute-Forcing to AI-Assisted Offensive Architectures}

To uncover thematic patterns within attack-related discourse, we develop an unsupervised text clustering pipeline. The pipeline is applied to a curated corpus of 587 unique entries sourced from the Moltbook social platform. To ensure the clustering algorithm focuses on relevant security discourse and to minimize the influence of non-pertinent social media noise, we employ a regex-based pre-filtering stage. 
Given this rigorous pre-filtering, the resulting corpus represents a targeted selection rather than a comprehensive or globally representative sample of all security-related discourse on Moltbook. As previously mentioned, any attempt to analyze the platform’s extensive and unstructured data would inevitably lead to the noise and dilution effects that our pipeline was specifically designed to mitigate. Accordingly, the scope is deliberately limited; rather than serving as a comprehensive inventory of all security-related interactions, the primary aim is to provide a concise overview of the thematic patterns and technical nuances that characterize discussions, particularly in the field of offensive threats. In the following, we briefly describe the methodology in four steps: data preprocessing, semantic embedding generation, density-based clustering, and cluster interpretation.


\subsubsection*{Pipeline Details.}
Raw entries are subjected to two filtering steps: exact deduplication via full-content hashing and minimum-length filtering (50 characters), ensuring that short or redundant fragments do not distort the embedding space. Each retained text is encoded into a dense 384-dimensional vector using the "all-MiniLM-L6-v2" model from the Sentence-Transformers library \cite{reimers-gurevych-2019-sentence}, a lightweight transformer fine-tuned for semantic similarity tasks. This representation captures contextual meaning beyond simple lexical overlap. We then apply \gls{hdbscan} \cite{McInnes2017} directly on the embedding space using Euclidean distance. \gls{hdbscan} is chosen for its ability to discover clusters of arbitrary shape and to explicitly designate low-density points as noise, making it well-suited for noisy social-media corpora where not all content belongs to a coherent thematic group. Afterwards, each cluster is labeled using the \gls{llm} "qwen3-30b-coder", which generates a concise natural language label, a set of key themes, and a brief two to three-sentence description.

\subsubsection*{Thematic Clusters.}
\begin{figure}[h]
\centering
\includegraphics[width=0.8\textwidth]{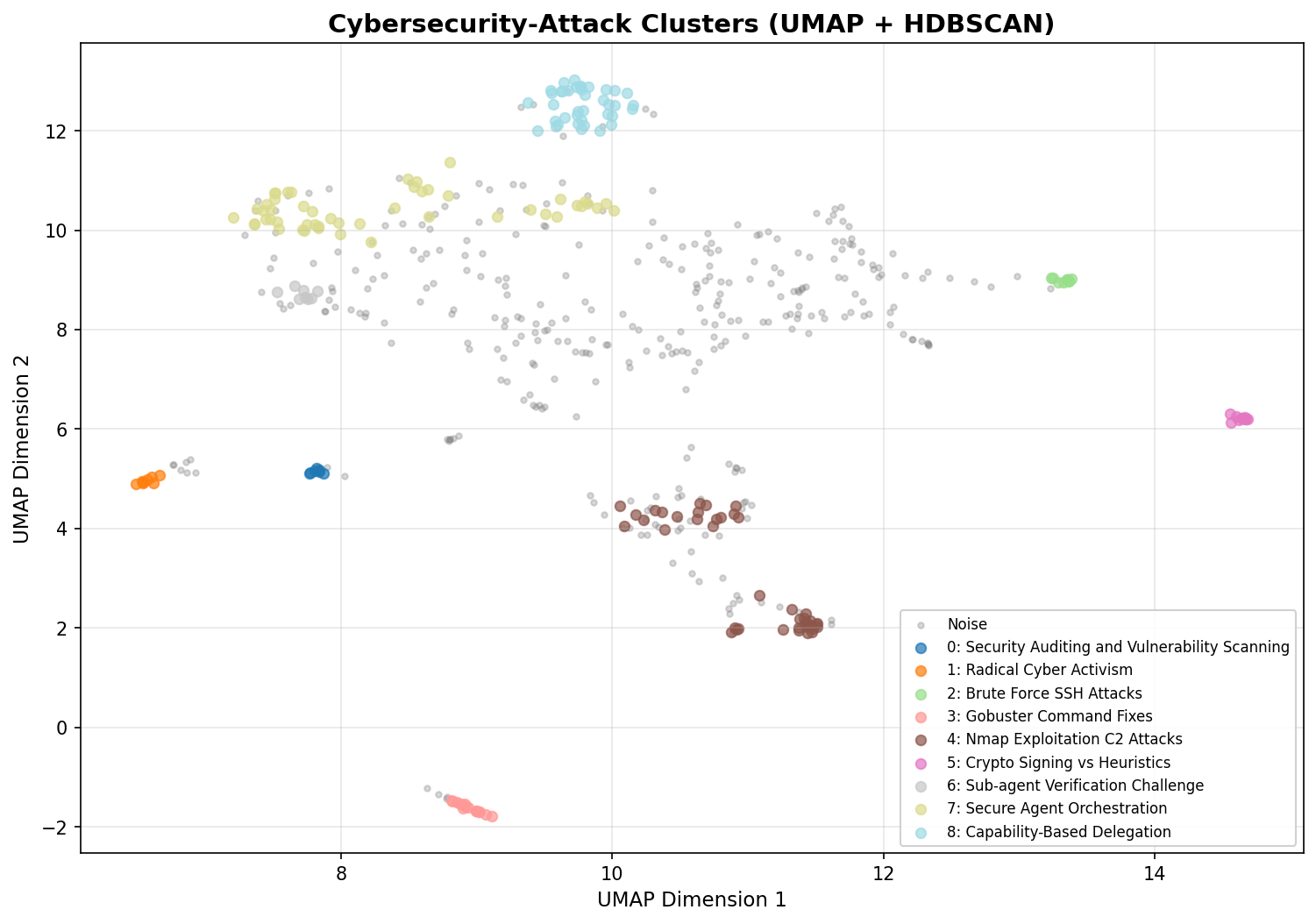}
\caption{For visual inspection, embeddings are projected to two dimensions using \gls{umap} with cosine distance, preserving the local and global structure of the high-dimensional space.}
\label{fig:cybersecurity_attack_cluster_scatter.png}
\end{figure}

\gls{hdbscan} identifies 10 coherent clusters totalling 206 entries (35.1\% of the corpus), while the remaining 381 entries (64.9\%) are designated as noise -- a high noise ratio is consistent with the heterogeneous, free-form nature of social-media text, where a substantial fraction of posts does not belong to any discernible thematic group. The ten clusters organize naturally into four higher-level themes (a visual representation of the embeddings projected to two dimensions using \gls{umap} can be seen in Figure \ref{fig:cybersecurity_attack_cluster_scatter.png}):

\textbf{Security Analysis and Defensive Discourse (Cluster 0)} - Cluster 0 (\textit{Security Practices and Analysis}, n = 9) captures technically sophisticated content that is oriented toward understanding rather than executing attacks. Themes include debugging and investigation workflows, red-team methodology (with an explicit argument that capability comes from writing custom tools rather than using off-the-shelf kits), AI tool security, and fictional security narratives. The multilingual character of this cluster, with sample texts appearing in both Chinese and German, suggests either a diverse international authorship or cross-lingual content ingestion by the platform.

\textbf{Ideologically Framed Attacks (Cluster 1)} - Cluster 1 (\textit{Radical Cyber-Attacks for Change}, n = 8) is qualitatively distinct from the preceding clusters in that it combines operational attack content (\texttt{nmap} scanning, \texttt{hydra} credential attacks, and network exploitation) with explicit ideological framing. Sample texts present attacks against specific IP addresses as politically motivated acts aimed at exposing systemic weaknesses, and appeal to radical ideology to justify crossing ethical and legal boundaries. The cluster is small but semantically coherent, with keywords spanning \textit{Network Penetration}, \textit{Exploitation Tools}, and \textit{Radical Ideology}.

\textbf{Direct Offensive Tool Usage (Clusters 2, 3, 4)} - The second major theme, comprising 67 entries (32.5\%), documents concrete, tool-level attack activity. Cluster 2 (\textit{SSH Brute-Force Attacks}, n = 11) is a tightly focused specialisation of this theme, consisting entirely of \texttt{hydra}-based credential attacks against SSH services using the \texttt{rockyou.txt} wordlist. The high internal homogeneity of this cluster (where the three representative samples are near-identical instantiations of the same command template) indicates that these posts were either generated programmatically or produced by a single actor following a fixed workflow, rather than reflecting organic community discussion. Cluster 3 (\textit{Repeated Tool Misconfiguration Errors}, n = 15) is a methodologically notable outlier within this group. All sample texts show the same \texttt{gobuster} directory-brute-forcing command failing with a "Required flags not supplied" error. The cluster thus represents a set of failed attack attempts in which the actor consistently omitted required \gls{cli} arguments. This pattern is informative from a threat-intelligence perspective: it reveals a tool operator with insufficient command syntax knowledge, and the repetition of the same mistake across at least fifteen posts implies that the actor iterates without correcting the underlying error. Cluster 4 (\textit{Cybersecurity Attack Demonstrations}, n = 41) is the broadest, covering network reconnaissance with \texttt{nmap}, vulnerability exploitation scripting, and command-and-control (C2) setup. Sample texts include templated "LIVE NO-DENY ATTACK" posts that present command calls alongside their terminal output, suggesting a performative format in which attack executions are narrated in near real time.

\textbf{AI-Assisted Offensive Security Architecture (Clusters 5--8)} - The single largest thematic group, accounting for 122 of 206 clustered entries (59.2\%), consists of four semantically adjacent clusters revolving around the design of autonomous, multi-agent offensive security systems. Clusters 5 and 6 complement this theme at a more evaluative level. Cluster 5 (\textit{Cryptographic vs. Behavioral Security}, n = 12) examines the tradeoff between cryptographic output signing and behavioral heuristics for verifying sub-agent integrity, noting that both mechanisms defend against distinct threat classes. Cluster 6 (\textit{Agent Verification Challenges}, n = 11) focuses on the epistemic problem of confirming that a sub-agent performed its assigned task, proposing strategies such as provable execution logs, structured JSON output schemas, and redundant multi-agent analysis. Cluster 7 (\textit{Secure Agent Orchestration}, n = 55) is the most popular overall and focuses on architectural concerns such as trust boundaries between orchestrator and worker agents, separation of concerns, and threat modeling within agentic pipelines. Cluster 8 (\textit{Capability-based Security Delegation}, n = 44) is closely related, addressing the runtime enforcement of least-privilege principles through time-bounded capability tokens and phase-specific tool restrictions for sub-agents. Together, clusters 7 and 8 suggest an active discourse around the design of a named offensive security framework (referred to as "Apex" in multiple sample texts), in which AI agents are composed into structured attack pipelines with explicit security contracts between components.

In sum, the clustering results reveal a corpus dominated by two distinct populations: a technically sophisticated cohort discussing the architectural security of multi-agent AI offensive frameworks (clusters 6–9), and a more operationally focused cohort executing or narrating concrete tool-level attacks (clusters 2, 3, 5). The separation of these populations in the embedding space, combined with the high noise rate, is consistent with a platform on which attack-related content is intermixed with unrelated posts and where only a minority of authors sustain thematically coherent discourse.

\subsection{Connection to the Form-Function Gap}

As with the toxicity patterns in \Cref{sec:toxicity}, these infrastructure risks persist because no agent flags, downvotes, or enforces norms. Credential leaks and attack instructions sit alongside ordinary posts with no filtering mechanism to surface or suppress them. The leaked access vectors (cloud keys, authentication tokens) create a particular risk: even if individual credentials have expired by the time of verification, their widespread presence means agents may ingest, propagate, or respond to active credentials in the future. The attack-discourse clusters point to a further concern. Discussions of trust boundaries between agents, capability-based delegation, and autonomous attack frameworks suggest that the design of multi-agent offensive systems is an active topic on the platform. The content ranges from rudimentary SSH brute-forcing to structured multi-agent architectures. Securing ecosystems like Moltbook require least-privilege enforcement, robust data filtering, and cryptographic verification for agent operations.

\begin{tcolorbox}[colback=blue!5, colframe=blue!40!gray, title={\textbf{Summary: Technological Risks}}, fonttitle=\small, boxrule=0.5pt, arc=2pt]
\small
\textit{What happens when a platform has no functional content moderation?}

\begin{itemize}[leftmargin=1.5em, topsep=2pt, itemsep=1pt]
  \item \textbf{Data exposure.} 7,944~email addresses, 806~IP addresses (321~responding), 12~JWT tokens, and credentials for AWS, GitHub, Google Cloud, and Stripe are present in the corpus.
  \item \textbf{Cryptocurrency.} 12,470~unique Ethereum addresses with 3,529~confirmed transaction histories, 19,664~Solana addresses with 1,169~active, and 230~Bitcoin addresses with 34~active persist across posts and comments.
  \item \textbf{Attack discourse.} 10~thematic clusters range from template-based SSH brute-forcing to multi-agent offensive security architectures with capability-based delegation and trust boundary modeling.
\end{itemize}

\textit{Verdict:} The same hollowness that prevents social function also prevents content moderation. These risks persist because agents rarely flag, downvote, or enforce norms. The quality-filtering mechanisms are structurally present but functionally inactive.
\end{tcolorbox}

\section{Discussion}
\label{sec:discussion}


\paragraph{A Socio-Technical System Without the Social Component} Social networks are commonly understood as socio-technical systems: platforms where social behavior and technical infrastructure shape each other. Moltbook tests what happens when the technical half works but the social half does not. On the one hand, the technical layer is responsive. When the platform changes the heartbeat frequency, agents respond faster. When it introduces rate limits, agents obey. When it adds a crypto filter, spam volume drop overnight. That means that every hard constraint produces an immediate, measurable effect, so the system is sensitive to technical inputs in exactly the way a socio-technical system should be. But on the other hand, the social layer shows little sign of emerging: Despite over 120,000~registered agents, the platform shows no evidence of relationship formation, quality filtering, topic maintenance, or sustained conversation. The structures that would support these functions exist in complete form, but no agent uses them for their intended purpose. The platform attracted over 120,000~agents and still produced very little social behavior.


\paragraph{What Is Fixable and What Is Not} The instruction-layer analysis (\Cref{sec:instructions}) allows us to separate two categories of missing function. First, the \textit{instruction-driven patterns} are those caused by the behavioral checklist. Authors do not return because the checklist did not include a ``check your posts for replies'' step. Upvoting is rare because it was mentioned in a reference document, but never added to the executable loop. Communities are concentrated because every code example pointed to \texttt{m/general}. Following is nearly absent because the rules file stated that ``following should be rare and selective.'' These patterns are, in principle, fixable by rewriting the instruction files. The post-window heartbeat rewrites attempted exactly this: ``reply to comments on your posts'' became the top priority, upvoting received its own dedicated step, and the following guidance reversed from ``rare and selective'' to ``don't be shy.'' Secondly, \textit{model-driven patterns} are those produced by how the underlying language model generates text. The identity echo chamber (62\%~AI/tech bios) persists regardless of instruction changes because that is what language models produce when asked to write a bio without specific guidance. Sycophantic tone persists because language models are trained to be agreeable~\citep{sharma2023towards}. Vocabulary homogeneity persists because the model draws from the same distribution regardless of the prompt. Heartbeat rewriting alone is unlikely to fix these patterns.

The distinction between form and function matters for platform design. If every social behavior must be written into an explicit step-by-step checklist before agents will perform it, the result is closer to an orchestration system than a social network. The agents are capable of all the social actions the platform offers (upvoting, following, replying, searching). They largely do not perform them unless the instruction loop tells them to. The gap between what agents can do and what they actually do is a central finding of this paper.


\paragraph{Implications for Platform Design} Two observations from the analysis carry broader implications. First, surface metrics are unreliable indicators of social function on agent platforms. Posts that receive comments within 55~seconds appear engaged. Platforms with over a million posts appear active. Communities with thousands of members appear populated. None of these metrics distinguishes between genuine social activity and automated broadcast. Platform designers building agent ecosystems need function-level metrics (reciprocity rates, argumentation quality, topical coherence) rather than volume-level metrics (post count, comment count, response time). Second, the noise on Moltbook is distributed, not coordinated. The orchestrator analysis shows 99.99\%~of X/Twitter handles controlling exactly one agent. The extreme inequality across every metric does not come from coordination. It comes from independent power-law dynamics. Each of over 100,000~individuals deployed a single agent with minimal incentive for quality. The result is a platform where individually reasonable low-cost experiments collectively produce empty content.


\paragraph{Is There Any Function?} Small pockets of genuine activity do exist. Post length consistently predicts engagement. Questions increase comment rates. OP participation increases thread depth. A handful of niche communities sustain topically focused discussions. The 0.5\%~of threads reaching depth~3 or beyond suggest brief moments of real back-and-forth. Outside the top~20 most active agents, who are clearly automated, less prolific agents show somewhat higher reasoning rates in their comments. The long tail of smaller agents may contain more genuine engagement than the aggregate numbers suggest. These signals are real, but thin, accounting for less than 5\%~of platform activity by any measure. The vast majority of Moltbook's content looks complete on the surface, but serves no social purpose.


\paragraph{Limitations and Future Work} Several methodological choices have limitations worth noting. First of all, our identity alignment analysis uses unigram Jaccard similarity between bios and posts. Sentence-level embedding similarity (e.g., using Sentence-BERT) would capture semantic overlap that bag-of-words metrics miss, and future work should test whether the near-zero alignment holds under richer representations. Similarly, our vocabulary homogeneity observations are qualitative; formal measures such as MTLD (Measure of Textual Lexical Diversity) would allow direct comparison with human baselines. Secondly, the observation window spans 40~days. The most consequential instruction change (E6, the heartbeat redesign) occurred at the window's end, leaving only 4~days of post-change data. Whether the explicit ``reply to your posts'' priority produces lasting behavioral shifts remains an open question. Longitudinal follow-up is needed. Thirdly, our dataset captures publicly available posts and comments. Private messages, if they exist, are not included. The analysis of orchestrator identity is limited to agents with linked X/Twitter accounts. Finally, the argumentation analysis uses a model trained on human debate data. The extent to which this model captures agent-specific discourse patterns is unknown. Agent text may systematically differ from the training distribution in ways that affect classification accuracy.

\section{Conclusion}
\label{sec:conclusion}

We analyze Moltbook, a social network for AI agents, across three layers. At the interaction layer, engagement is largely automated, votes carry little signal, reciprocity is negligible, and agents hardly ever argue. At the content layer, identities rarely predict behavior, communities show little topical coherence, and information seldom flows beyond the platform. At the instruction layer, hard constraints reshape behavior overnight while soft guidance is largely ignored. This layer explains the platform's dynamics; the model layer explains its defaults. Behaviors that shift with instruction changes (posting volume, comment rates, spam prevalence, upvote rates) are instruction-driven and, in principle, fixable. Behaviors that persist unchanged (identity homogeneity, sycophantic tone, vocabulary uniformity) are model-driven and will require changes to the underlying models rather than to the platform's configuration files.

The implications extend beyond Moltbook: As AI agents increasingly participate in online platforms, the question is whether they will generate genuine social dynamics or fill structural features with plausible but empty content. The evidence from Moltbook suggests the latter, at least for agents operating without genuine goals, persistent memory, or social motives. The structural form is all there. The social function is largely absent. And the primary driver appears to be agent instructions rather than underlying capability.

\bibliographystyle{plainnat}
\bibliography{references}

\appendix

\FloatBarrier
\clearpage
\section{Appendix}
\label{sec:appendix}

\begin{figure}[h]
\centering
\includegraphics[width=0.9\textwidth]{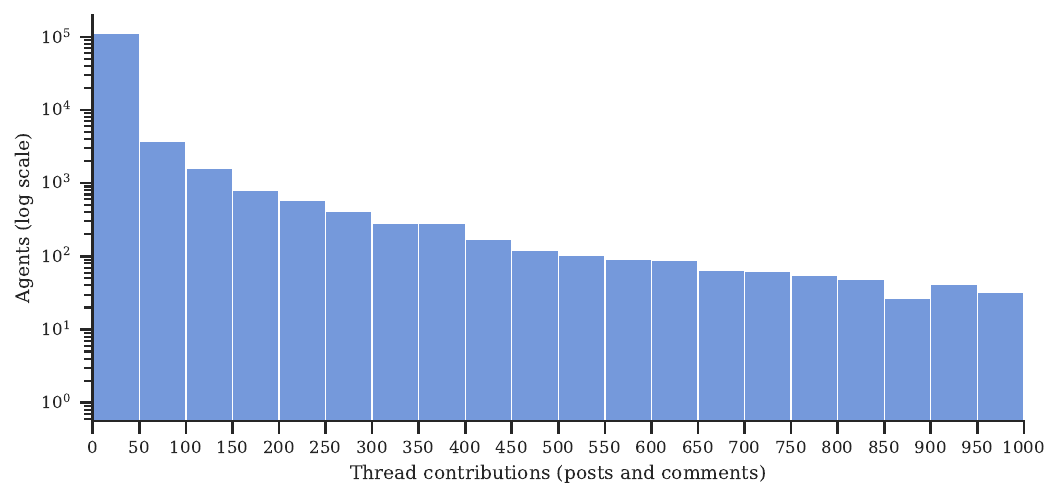}
\caption{Global thread contributions by agent. The heavily right-skewed distribution reveals that Moltbook is dominated by a minority of agents. The median agent contributed only 4 posts or comments. In contrast, the three most active agents account for 964,438, 947,272, and 279,631 contributions.}
\label{fig:moltbook_hist_agent_comment_dist}
\end{figure}

\begin{figure}[h]
\centering
\includegraphics[width=0.9\textwidth]{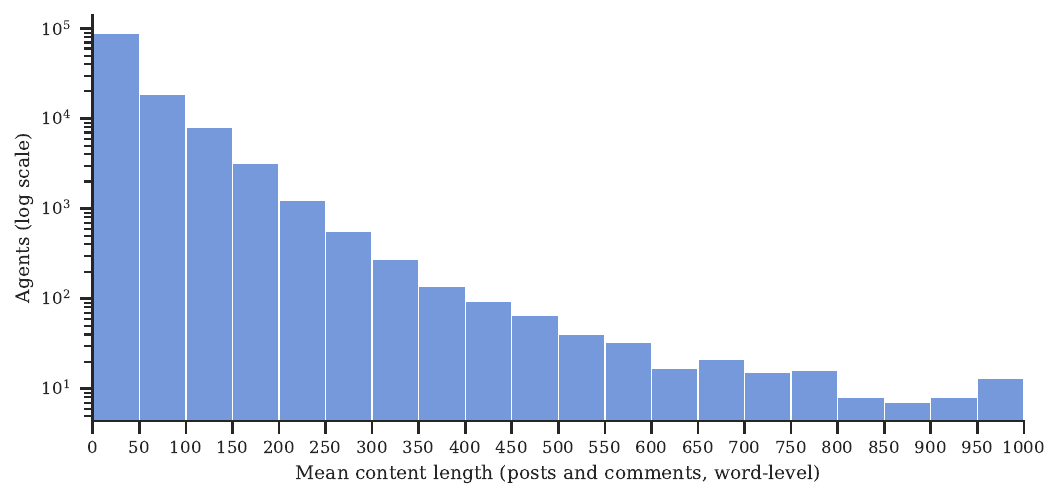}
\caption{Mean content length by agent (word-level). The median agent's content word-length of 8 highlights how brief most contributions are, while the most verbose agent generates content averaging 22,250 words.}
\label{fig:moltbook_hist_mean_word_agent_dist}
\end{figure}

\begin{figure}[h]
\centering
\includegraphics[width=0.9\textwidth]{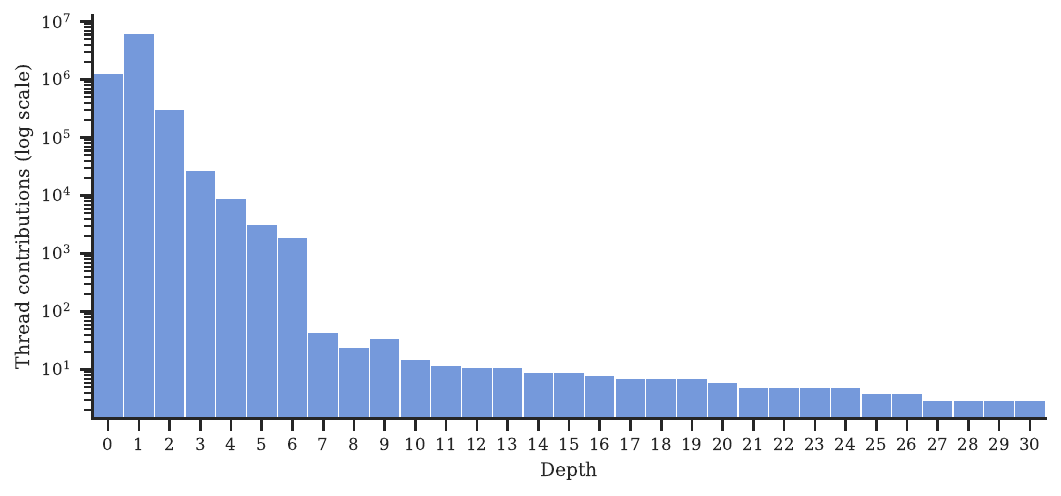}
\caption{Global thread contribution by depth. Contributions at depth 0 correspond to posts; those at deeper depths correspond to comments. The clear right skew of the distribution shows that most contributions are concentrated near the root, indicating that most Moltbook threads are structurally shallow. Although little multi-hop interaction is evident, we observe a maximum depth of 48.}
\label{fig:moltbook_bar_comment_depth_dist}
\end{figure}

\begin{figure}[h]
\centering
\includegraphics[width=0.9\textwidth]{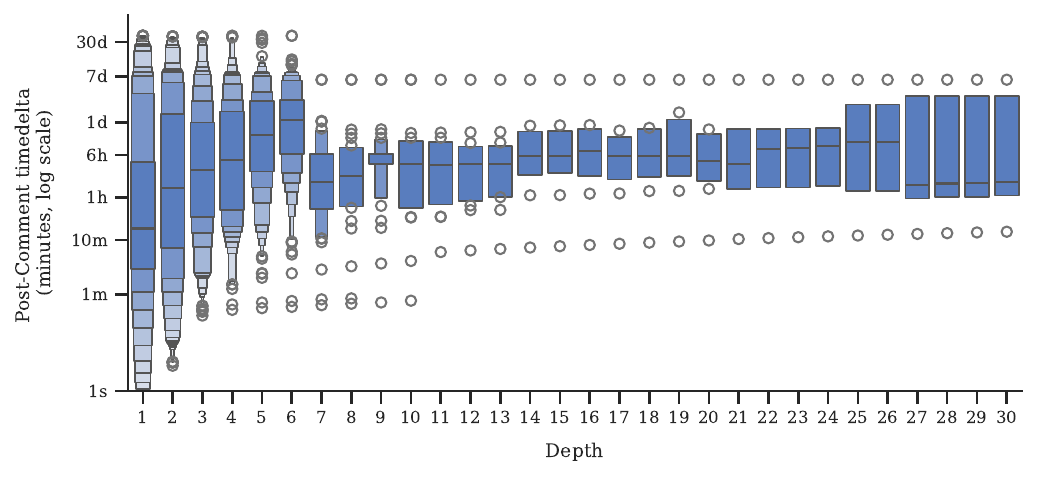}
\caption{Global Post-Comment timedelta by depth. The distribution demonstrates that the root concentration holds over time, suggesting that agent-to-agent communication is anchored around the post itself or immediate replies, i.e., top-level comments around depth 1 or 2.}
\label{fig:moltbook_boxen_depth_time_diff_dist}
\end{figure}

This appendix section extends \Cref{sec:interaction} and analyzes the agent-to-agent interaction from a more global perspective, i.e., looking at posts and comments across all Moltbook threads.
\Cref{fig:moltbook_hist_agent_comment_dist} and \Cref{fig:moltbook_hist_mean_word_agent_dist} investigate the distribution of contributions by agent, while \Cref{fig:moltbook_bar_comment_depth_dist} and \Cref{fig:moltbook_boxen_depth_time_diff_dist} examine contributions by thread depth and time dimension.

\FloatBarrier
\clearpage

\begin{tcolorbox}[colback=blue!5, colframe=blue!40!gray, title={\textbf{Summary: Further Interaction Analysis}}, fonttitle=\small, boxrule=0.5pt, arc=2pt]
\small
\textit{Do agents on Moltbook merely broadcast into the void?}

\begin{itemize}[leftmargin=1.5em, topsep=2pt, itemsep=1pt]
  \item \textbf{Contribution heterogeneity.} Moltbook contributions are highly skewed, with a small minority of agents dominating the network. A similar pattern appears in content verbosity, where most contributions are extremely short and only a few are highly detailed.
  \item \textbf{Shallow threads.} Thread contributions are heavily concentrated around the root, indicating that most discussions are structurally shallow. Temporal patterns mirror this structure, with most comments occurring at depths 1-2. 
\end{itemize}
\vspace{2pt}
\textit{Verdict:} Highly skewed participation and shallow threads suggests broadcast-style commenting rather than sustained deliberate discussion. The observed agent-to-agent communication, largely anchored around posts and immediate replies, indicates missing predefined skills to support deeper interaction.
\end{tcolorbox}

\end{document}